%% file: Main-part.tex
\documentclass[lettersize,journal]{IEEEtran}
\usepackage{amsmath,amsfonts}
\usepackage{array}
\usepackage[caption=false,font=normalsize,labelfont=sf,textfont=sf]{subfig}
\usepackage{textcomp}
\usepackage{stfloats}
\usepackage{url}
\usepackage{verbatim}
\usepackage{graphicx}

\usepackage{cite}
\usepackage{amsmath,amssymb,amsfonts}
\usepackage{algorithm,algorithmic}
\usepackage{textcomp}
\usepackage{amsmath}  
\usepackage{amsfonts}  
\usepackage{amssymb}  
\usepackage{soul}
\usepackage{enumitem}
\usepackage{xcolor}
\usepackage{booktabs}
\usepackage{cuted}
\usepackage{svg}
\usepackage{tcolorbox}
\usepackage{bm}

\def\BibTeX{{\rm B\kern-.05em{\sc i\kern-.025em b}\kern-.08em
    T\kern-.1667em\lower.7ex\hbox{E}\kern-.125emX}}
\usepackage{balance}
\begin{document}
\title{Fluid Antenna-Aided Robust Secure Transmission for RSMA-ISAC Systems}
\author{Cixiao~Zhang,~\IEEEmembership{}
        Yin~Xu,~\IEEEmembership{Senior~Member,~IEEE,}
        Size~Peng,~\IEEEmembership{}
        Xinghao~Guo,~\IEEEmembership{}
        Xiaowu~Ou,~\IEEEmembership{Student~Member,~IEEE,}
        Hanjiang~Hong,~\IEEEmembership{Member,~IEEE,}
        Dazhi~He,~\IEEEmembership{Senior~Member,~IEEE,}
        and~Wenjun~Zhang,~\IEEEmembership{Fellow,~IEEE}
\thanks{This work was supported in part by the National Key Research and Development Project of China under Grant 2023YFF0904603; in part by the National Natural Science Foundation of China Program under Grant 62422111, Grant 62371291, and Grant 62271316. \textit{(Corresponding author: Yin Xu.)}

Cixiao Zhang, Size Peng, Xiaowu Ou, Xinghao Guo, Yin Xu, Dazhi He, and Wenjun Zhang are with the Cooperative Medianet Innovation Center, Shanghai Jiao Tong University, Shanghai 200240, China. Dazhi He is also affiliated with Pengcheng Laboratory, Shenzhen 518055, China (e-mail: \{cixiaozhang, sjtu2019psz, xiaowu\_ou, guoxinghao, xuyin, hedazhi, zhangwenjun\}@sjtu.edu.cn). 

Hanjiang Hong is with the Department of Electronic and Electrical Engineering, University College London, Torrington Place, WC1E7JE, United Kingdom (e-mail: hanjiang.hong@ucl.ac.uk). }}

\markboth{Journal of \LaTeX\ Class Files,~Vol.~18, No.~9, September~2020}%
{How to Use the IEEEtran \LaTeX \ Templates}

\maketitle

\begin{abstract}
This paper leverages fluid antenna (FA) and rate-splitting multiple access (RSMA) to enhance the physical layer security (PLS) of an integrated sensing and communication (ISAC) system. We consider a practical multi-user multi-input single-output (MU-MISO) system, where a base station (BS) equipped with fixed position antennas (FPAs) employs RSMA to communicate with multiple single-FA users, while an eavesdropping target may potentially wiretap the signals. The system adopts a novel rate splitting (RS) scheme, where the common layer stream serves a dual purpose: it conveys valid data to legitimate users (LUs) while simultaneously generating jamming signals to confuse potential eavesdroppers. We establish the problem and propose the optimization algorithm under two conditions: perfect and imperfect channel state information (CSI) conditions. Specifically, under perfect the CSI condition, we address the non-convex optimization problem by proposing an alternating optimization (AO) algorithm, which decomposes the problem into two subproblems: beamforming matrix optimization and the adjustment of FA positions. For beamforming optimization, we utilize semidefinite programming (SDP) and successive convex approximation (SCA) to convert the problem into a more tractable convex form. Given a fixed beamforming matrix, SCA is applied to handle the surrogate upper bound of the constraints. In the case of imperfect CSI, the continuous nature of CSI errors leads to an infinite number of constraints. To overcome this challenge, we propose an AO-based algorithm that incorporates the S-Procedure and SCA to obtain a high-quality beamforming matrix and effective FA positions. Extensive simulation results demonstrate that the proposed FA-aided RSMA-ISAC system significantly enhances security compared to traditional FPA-based and SDMA-based systems.
\end{abstract}


\input{Chapter/intro-v2}
\input{Chapter/system_model}

\input{Chapter/proposed-solution}

\input{Chapter/simulation}
\input{Chapter/conclusion}
\appendix
\section{Appendix A}
\subsection{Derivations of $\nabla J(\mathbf{p}_k)$ and $\delta_{k,1}$ in (\ref{J upper})}
\label{Appendix perfect antenna 1}
 The gradient vector of $J(\mathbf{p}_k)$ is

\begin{equation}
\scalebox{0.74}{$
\nabla J(\mathbf{p}^{(t)}_k) = 
\begin{bmatrix} 
\frac{4\pi
}{\lambda}\sum\limits_{l=1}^{L_k}\mathbf{q}_{k,1}(l)\sin\left[-\frac{2\pi}{\lambda}\rho^r_{k,l}(\mathbf{p}_k^{(t)}+\angle \mathbf{q}_{k,1}(l))\right]\sin{\theta_l}\cos{\phi_l} \\ 
\frac{4\pi
}{\lambda}\sum\limits_{l=1}^{L_k}\mathbf{q}_{k,1}(l)\sin\left[-\frac{2\pi}{\lambda}\rho^r_{k,l}(\mathbf{p}_k^{(t)}+\angle \mathbf{q}_{k,1}(l))\right]\cos{\theta_l}
\end{bmatrix},
$}
\end{equation}
Given the inequality
\begin{equation}
||\nabla J(\mathbf{p}_k)||2^2 \leq ||\nabla J(\mathbf{p}_k)||F^2 \leq 4\left(\frac{8\pi^2}{\lambda^2}\sum\limits_{l=1}^{L_k}|\mathbf{q}_{k,1}(l)|\right)^2,
\end{equation}
we can select
\begin{equation}
\delta_{k,1} = \frac{16\pi^2}{\lambda^2}\sum\limits_{l=1}^{L_k}|\mathbf{q}_{k,1}(l)|.
\end{equation}

\subsection{Derivations of $\nabla T_k(\mathbf{p}_k)$ and $\delta_{k,2}$ in (\ref{T upper})}
\label{Appendix T upper}
Similarly, we have
\begin{equation}
    \scalebox{0.75}{$
    \nabla T_k(\mathbf{p}^{(t)}_k)=
    \begin{bmatrix} 
    \frac{4\pi
    }{\lambda}\sum\limits_{l=1}^{L_k}\mathbf{q}_{k,2}(l)\sin[-\frac{2\pi}{\lambda}\rho^r_{k,l}(\mathbf{p}_k^{(t)}+\angle \mathbf{q}_{k,2}(l)]\sin{\theta_l}\cos{\phi_l} \\ 
    \frac{4\pi
    }{\lambda}\sum\limits_{l=1}^{L_k}\mathbf{q}_{k,2}(l)\sin[-\frac{2\pi}{\lambda}\rho^r_{k,l}(\mathbf{p}_k^{(t)}+\angle \mathbf{q}_{k,2}(l)]\cos{\theta_l}
    \end{bmatrix},
    $}
\end{equation}
\begin{equation}
    \delta_{k,2}=\frac{16\pi^2}{\lambda^2}\sum\limits_{l=1}^{L_k}|\mathbf{q}_{k,2}(l)|.
\end{equation}
\subsection{Derivations of $\nabla D_k(\mathbf{p}_k)$ and $\delta_{k,3}$ in (\ref{D upper})}
\label{Appendix D upper}
\begin{equation}
\nabla D_k(\mathbf{p}_k^{(t)}) = \frac{4\pi}{\lambda}
\scalebox{0.75}[0.75]{\ensuremath{  
\begin{bmatrix}
    \sum\limits_{l=1}^{L_k} |\boldsymbol{a}_k(l)| \sin[ -\frac{2\pi}{\lambda}\rho^r_{k,l}(\mathbf{p}_k^{(t)}) + \angle \boldsymbol{a}_k(l) ] \sin\theta_l \cos\phi_l \\
    \sum\limits_{l=1}^{L_k} |\boldsymbol{a}_k(l)| \sin[ -\frac{2\pi}{\lambda}\rho^r_{k,l}(\mathbf{p}_k^{(t)}) + \angle \boldsymbol{a}_k(l) ] \cos\theta_l
\end{bmatrix},
}}
\label{eq:gradient_scaled}
\end{equation}
\begin{equation}
    \delta_{k,3}=\frac{16\pi^2}{\lambda^2}\sum\limits_{l=1}^{L_k}|\boldsymbol{a}_k^{(t)}|.
\end{equation}

\subsection{Derivations of $\nabla_x \Gamma_1(\mathbf{p}_k)$, $\nabla_y \Gamma_1(\mathbf{p}_k)$ and $\delta_{k,4}$ in (\ref{Gamma1})}
\label{Appendix gammas}
Let 
\begin{equation}
\scalebox{0.90}{$
    \boldsymbol{c}^{(t)}_k=[\sum\limits_{l=1}^{L_k}e^{-j\frac{2\pi}{\lambda}\rho^{r,(t)}_{k,l}}\mathbf{\Pi}_{k,1}[l,1],\dots,\sum\limits_{l=1}^{L_k}e^{-j\frac{2\pi}{\lambda}\rho^{r,(t)}_{k,l}}\mathbf{\Pi}_{k,1}[l,N_T] ]^\top,
    $}
\end{equation}
\begin{align}
    &\nabla_x \Gamma_1(\mathbf{p}_k^{(t)})=2\Re\{(\nabla_x\boldsymbol{c}_k^{(t)})^\top\Delta \mathbf{h}\}\nonumber\\
    &=(\nabla_x\boldsymbol{c}_k^{(t)})^\top\Delta \mathbf{h}+\Delta\mathbf{h}^H(\nabla_x \boldsymbol{c}^{(t)}_k)^*,\\
    &\nabla_y \Gamma_1(\mathbf{p}_k^{(t)})=2\Re\{(\nabla_y\boldsymbol{c}_k^{(t)})^\top\Delta \mathbf{h}\}\nonumber\\
    &=(\nabla_y\boldsymbol{c}_k^{(t)})^\top\Delta \mathbf{h}+\Delta\mathbf{h}^H(\nabla_y \boldsymbol{c}_k^{(t)})^*.
\end{align}
Thus, we have
\begin{equation}
    \delta_{k,3}=\frac{16\pi^2}{\lambda^2}\sum\limits_{l=1}^{l_k}\epsilon_k\sqrt{\boldsymbol{\Pi}_{k,1}[l,:]^H\boldsymbol{\Pi}_{k,1}[l,:]}.
\end{equation}
\subsection{Derivations of $\nabla C(\mathbf{p}_k^{(t)})$ and $\delta_{k,5}$ in (\ref{Gamma2 part2})}
\label{Appendix Gamma2 part2}

Assume
\begin{equation}
   \mathbf{q}_{k,2}=-(\mathbf{\Lambda}_{k,2}-\mathbf{\Pi}_{k,2})\mathbf{f}_k(\mathbf{p}_k^{(t)}),
\end{equation}
So we can get
\begin{equation}
\scalebox{0.82}{$
    \nabla C(\mathbf{p}_k^{(t)})=
    \begin{bmatrix} 
    \frac{4\pi
    }{\lambda}\sum\limits_{l=1}^{L_k}(|\mathbf{q}_{k,2}(l)|\sin[-\frac{2\pi}{\lambda}\rho^r_{k,l}(\mathbf{p}_k^{(t)}+\angle \mathbf{q}_{k,2}(l)]\sin{\theta_l}\cos{\phi_l} \\ 
    \frac{4\pi
    }{\lambda}\sum\limits_{l=1}^{L_k}|\mathbf{q}_{k,2}(l)|\sin[-\frac{2\pi}{\lambda}\rho^r_{k,l}(\mathbf{p}_k^{(t)}+\angle \mathbf{q}_{k,2}(l)]\cos{\theta_l}
    \end{bmatrix},
    $}
\end{equation}
\begin{equation}
    \delta_{k,6}=\frac{16\pi^2}{\lambda^2}\sum\limits_{l=1}^{L_k}|\mathbf{q}_{k,2}(l)|.
\end{equation}
\subsection{Derivations of $\mathbf{z}_{k,2}$ and $\delta_{k,6}$ in (\ref{Uk,1 uppper})}
\label{Appendix Uk1}
Assume
\begin{equation}
\scalebox{0.88}{$
    [\nabla_xU_{k,1} ~\nabla_yU_{k,1}](\mathbf{p}_k-\mathbf{p}_k^{(t)})=2\Re\{\Delta \mathbf{h}^H[\nabla_x \mathbf{t}_k^* ~\nabla_y \mathbf{t}_k^*](\mathbf{p}_k-\mathbf{p}_k^{(t)})\},
$}
\end{equation}
\begin{equation}
    \mathbf{z}_{k,2}=[\nabla_x \mathbf{t}_k^* ~\nabla_y \mathbf{t}_k^*](\mathbf{p}_k-\mathbf{p}_k^{(t)}),
\end{equation}
where
\begin{equation}
    \nabla_x \mathbf{t}_k=[-j\frac{2\pi}{\lambda}\sum\limits_{l=1}^{L_k}(\sin\theta_{k,l}\cos\phi_{k,l})e^{-j\frac{2\pi}{\lambda}\rho^{(t)}_{k,l}}\mathbf{B}[l,1]]^{l=L\times 1},
\end{equation}
\begin{equation}
    \nabla_y \mathbf{t}_k=[-j\frac{2\pi}{\lambda}\sum\limits_{l=1}^{L_k}(\cos\theta_{k,l})e^{-j\frac{2\pi}{\lambda}\rho^{(t)}_{k,l}}\mathbf{B}[l,1]]^{l=L\times 1}.
\end{equation}
\begin{equation}
    \delta_{k,6}=\frac{16\pi^2}{\lambda^2}\sum\limits_{l=1}^{l_k}\epsilon_k\sqrt{\mathbf{B}_k[l,:]^H\mathbf{B}_k[l,:]}.
\end{equation}
\subsection{Derivations of $\nabla U_{k,2}(\mathbf{p}_k^{(t)})$ and $\delta_{k,7}$ in (\ref{Uk2 upper bound})}
\label{Appendix uk2}
\begin{equation}
\scalebox{0.8}{$
    \nabla U_{k,2}(\mathbf{p}_k^{(t)})=    \begin{bmatrix} 
    \frac{4\pi
    }{\lambda}\sum\limits_{l=1}^{L_k}(|\mathbf{b}_k(l)|\sin[-\frac{2\pi}{\lambda}\rho^r_{k,l}(\mathbf{p}_k^{(t)}+\angle \mathbf{b}_k(l)]\sin{\theta_{k,l}}\cos{\phi_{k,l}} \\ 
    \frac{4\pi
    }{\lambda}\sum\limits_{l=1}^{L_k}|\mathbf{b}_k(l)|\sin[-\frac{2\pi}{\lambda}\rho^r_{k,l}(\mathbf{p}_k^{(t)}+\angle \mathbf{b}_k(l)]\cos{\theta_{k,l}}
    \end{bmatrix},
$}
\end{equation}
\begin{equation}
    \delta_{k,7}=\frac{16\pi^2}{\lambda^2}\sum\limits_{l=1}^{l_k}|\mathbf{b}_k(l)|.
\end{equation}

\bibliographystyle{IEEEtran}
\bibliography{IEEEabrv,refs}

\end{document}

%% file: Chapter/intro-v2.tex
\section{Introduction}
\IEEEPARstart{F}{luid} antenna (FA), a promising next-generation reconfigurable antenna (NGRA) technology, has been shown to enhance the flexibility and robustness of wireless communication systems \cite{Fasurvey1,Fasurvey2,Fasurvey3,Fasurvey4}. Recent advancements in FAs have inspired a variety of NGRA approaches, including movable antenna (MA), that are able to modify antenna properties such as position, shape, and orientation\cite{MAsurvey1, MAsurvey2, MAsurvey3}. Specifically, FAs dynamically adjust position within feasible regions or port activation, offering a transformative paradigm for improving communication, sensing, and other physical layer applications in wireless networks\cite{FAapplication1, FAapplication2, FAapplication3,honghan, Guo2024FluidAI}. Through flexible reconfigurations, FAs provide spatial degrees of freedom (DoFs), enabling the adjustment of phases across various propagation paths, which provides opportunities to enhance data transmission and defend against potential threats.

Recent works have increasingly integrated FA and other NGRA technologies with advanced physical layer techniques, particularly integrated sensing and communication (ISAC). ISAC allows sensing and communication functions to operate within the same system, which, combined with the widespread deployment of millimeter-wave and massive multiple-input multiple-output (MIMO) technologies, can enhance sensing accuracy with higher resolution in both time and angular domains. \cite{FAISAC1, FAISAC2zhou, FAISAC3Zhang,peng2024jointantennapositionbeamforming} focus on how novel NGRA methods can be applied to improve the performance of ISAC systems. Specifically, \cite{FAISAC1} and \cite{FAISAC2zhou} both address the trade-off problem in the FA-enabled ISAC system by jointly optimizing the transmit beamforming and port selection of FAs. \cite{FAISAC3Zhang} focuses on reducing the computational complexity of non-convex FA programming in the ISAC system. \cite{peng2024jointantennapositionbeamforming} uses MAs to reduce the self-interference of a general full-duplex ISAC scenario. However, these studies have not considered the practical security concerns of ISAC, as they assume that the target will not wiretap the information. In practice, there are severe scenarios, such as signal eavesdropping by an eavesdropping target. To guarantee physical layer security (PLS), many existing works \cite{FAsecure1,ma2024movable, MAsecure3} explore how to utilize spatial DoFs of NGRA to reduce risks. In particular, \cite{FAsecure1} considers a scenario in which a transmitter with a single fixed antenna position (FPA) sends confidential information to a legitimate receiver equipped with a planar FA array, while a potential eavesdropper (PE), also using a planar array, attempts to decode the message, revealing that using a fluid antenna with only one activated port can ensure more secure and reliable transmission. \cite{ma2024movable} introduces MAs and reconfigurable intelligent surfaces (RIS) in an ISAC system to maximize the sum rate of users while minimizing secrecy leakage. \cite{MAsecure3} examines an MA-aided secure MIMO communication system involving a BS, a legitimate receiver, and a PE, where the BS utilizes MAs to enhance the system's secrecy rate and extend the approach to a more general multicast case.

While the aforementioned strategies leverage spatial DoFs of NGRAs, they are all built upon space division multiple access (SDMA) systems. However, SDMA inherently limits performance due to several constraints, especially low spectral and energy efficiency, as well as restricted flexibility in handling highly correlated channels between the BS and users\cite{rsmasdma1,rsmasdma2}. Performance can be further improved by exploiting the DoFs of the streams provided by efficient multiple access techniques, enhancing PLS. Rate-Splitting Multiple Access (RSMA) has emerged as a promising next-generation multiple-access solution for many systems\cite{RSMAsurvey1,RSMAsurvey2,RSMAsurvey3}. RSMA fully exploits rate splitting, beamforming, and successive interference cancellation (SIC) by enabling partial interference decoding while treating the remaining interference as noise. In \cite{fu2020robust}, a method is proposed where the common layer stream in RSMA serves as a replacement for traditional artificial noise (AN). While AN increases the noise for PEs, it also interferes with legitimate users (LUs), as it does not contain useful information. The dual use of the common message in RSMA secure transmission systems serves two purposes: it acts as AN to confuse PEs while simultaneously being used as a useful message to improve the legitimate transmission sum rate. Using this property of RSMA significantly enhances the confidentiality of the communication. Recent studies have explored the integration of RSMA with ISAC security\cite{RSMAISACsecurity1, RSMAISACsecurity2}. \cite{RSMAISACsecurity1} obtains excellent security performance in an ISAC system by properly designing the beamformer. The authors of \cite{RSMAISACsecurity2} not only utilized RSMA but also employed RIS to establish a virtual line-of-sight link for target detection and to assist user communication, ultimately achieving significant improvements in security performance. However, these studies are based on traditional FPA configurations, so the spatial variation of the channel was not exploited.

There have been notable attempts to integrate NGRA with advanced multiple access techniques to enhance system performance. For instance, \cite{FAnoma1} investigates an FA-aided non-orthogonal multiple access (NOMA) system without channel state information (CSI) and demonstrates its superior performance over FPA configurations even with optimal CSI. Furthermore, \cite{RSMAFA1} presents the distribution of the equivalent channel gain and derives a compact expression for the outage probability, revealing that FA-aided RSMA systems outperform both FPA-based and NOMA-based schemes. Additionally, \cite{RSMAMA2} proposes an efficient two-stage coarse-to-fine-grained search (CFGS) strategy to optimize MA positions, significantly enhancing the sum rate of RSMA systems. These findings highlight the strong adaptability of FA and other NGRA technologies. 

Building on these advancements, it is noteworthy that the application of FAs in more complex but theoretically higher-performing RSMA-ISAC systems with eavesdropping targets for secure transmission remains underexplored. Moreover, in practical scenarios with channel estimation errors, this system becomes increasingly complex, and the design of FA/RSMA-aided secure transceivers cannot be directly applied without further investigation. Motivated by the above discussions and to fill this gap, this paper focuses on a downlink FA-aided RSMA-ISAC system to maximize the sum secure rates under both perfect and imperfect channel conditions. The main contributions of this paper are threefold.

\begin{itemize}
    \item We pioneer the study of the application of FAs in an RSMA-ISAC downlink system with perfect or imperfect CSI for both LUs and an eavesdropping target. A novel secure transmission RSMA strategy is adopted, where the common layer stream serves dual purposes: it conveys useful information for LUs and acts as AN for the eavesdropping target. Subsequently, we formulate an optimization problem aimed at maximizing the secrecy sum rate while ensuring the minimum sensing SNR constraint for the target. This is achieved by jointly optimizing the beamforming matrices and the positions of the FAs.

    \item We propose robust optimization algorithms for both the ideal case with perfect CSI and the practical case with imperfect CSI. When BS has perfect CSI, we first derive the secrecy sum rate upper bound through optimal FA positioning and beamforming design. Then, we develop a robust joint optimization framework to handle CSI uncertainties and maximize the worst-case secrecy rate via equivalent reformulation. Specifically, with perfect CSI, we tackle the non-convex optimization problem by proposing an AO algorithm, which decomposes it into two subproblems. For beamforming optimization, we use semidefinite programming (SDP) and SCA to convert the problem into a more tractable convex form. Given a fixed beamforming matrix, SCA is applied to address the surrogate upper bound of the constraints. In the case of imperfect CSI, the continuous nature of CSI errors introduces an infinite number of constraints. To overcome this, we propose an AO-based algorithm that combines the S-Procedure and SCA to obtain a high-quality beamforming matrix and effective antenna position adjustment.

    \item Numerical results demonstrate that the proposed FA-aided RSMA-ISAC system outperforms the FPA and SDMA schemes, achieving superior performance with our algorithm. Furthermore, our algorithm exhibits robustness across perfect and imperfect CSI, varying power budgets, and different configurations of users and antennas.

\end{itemize}

The rest of this paper is organized as follows. Section \ref{system model} presents the system model and problem formulation. In Section \ref{perfect CSI}, we propose the beamforming design and FA positioning under ideal conditions and discuss robust solutions for imperfect CSI in Section \ref{imperfect CSI}. Simulations are provided in Section \ref{simulation} to demonstrate the performance of the proposed solution. Finally, conclusions are drawn in Section \ref{conclusion}.

\textit{Notations:} Scalar variables are denoted by italic letters, vectors are denoted by boldface small letters, and matrices are denoted by boldface capital letters. $\mathbf{x}(n)$, $\mathbf{x}^\top$, $\mathbf{x}^{*}$, $\text{Tr}(\mathbf{X})$, $(\mathbf{X})^{-1}$, $\text{Rank}(\mathbf{X})$ , $\mathbf{X}[i,:]$ and $\mathbf{X}[j,i]$ denote the $n^{th}$ entry of $\mathbf{x}$, the transpose of $\mathbf{x}$, the conjugate of $\mathbf{x}$, the trace of $\mathbf{X}$, the inverse of $\mathbf{X}$ , the rank of $\mathbf{X}$, the $i^{th}$ line of $\mathbf{X}$ and the entry in the $j^{th}$ row and $i^{th}$ column of the matrix $\mathbf{X}$, respectively. The real and imaginary parts of a complex variable $\mathbf{X}$ are denoted by $\Re\{\mathbf{X}\}$ and $\Im\{\mathbf{X}\}$,respectively. $[m]^+$ represents $\max\{m,0\}$. $|\mathbf{n}|$ and $\angle \mathbf{n}$ denote the amplitude and phase of complex $\mathbf{n}$. 

%% file: Chapter/system_model.tex
\section{System Model}
\label{system model}
\begin{figure}[t]
    \centering
    \includegraphics[width=0.45\textwidth]{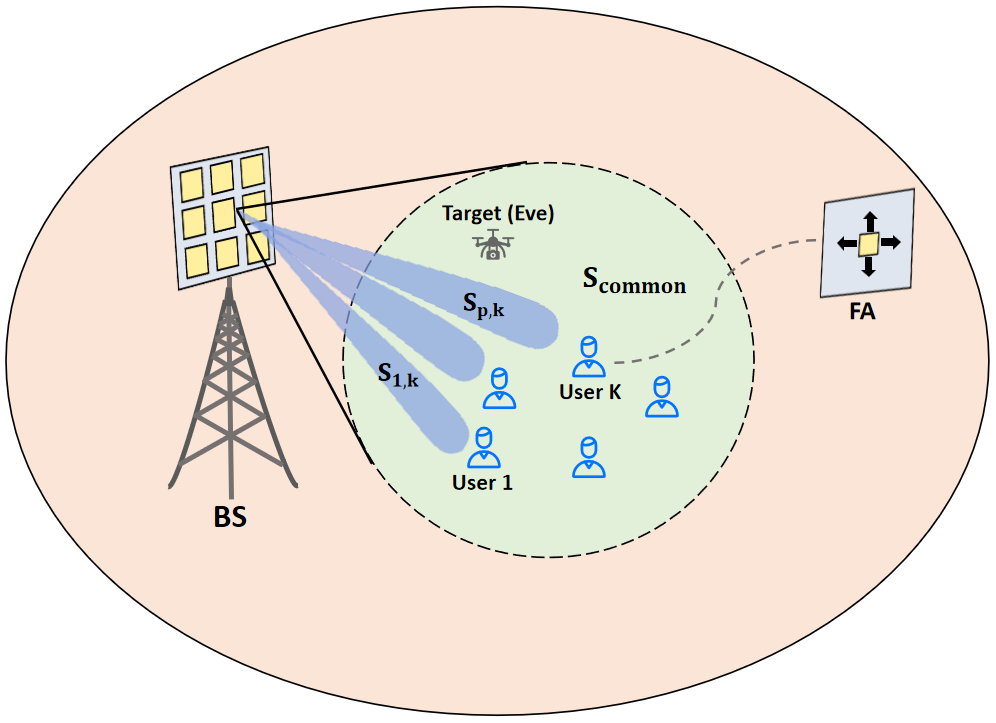}
    \caption{System model of the FA-aided RSMA-ISAC for secure transmission.}
    \label{fig: system model}
\end{figure}

Consider an RSMA-ISAC downlink scheme where a BS with a uniform planar array (UPA) of $N_T$ FPAs serves $K$ LUs, each equipped with a single FA, while an eavesdropping target, Eve, with a single FPA, attempts to intercept their data as shown in Fig. \ref{fig: system model}. The users are indexed by $\mathcal{K} = \{1, 2, \ldots, K\}$. 

\subsection{Channel Model}
Denote the antenna positions of the BS as $\mathbf{b}_n=[u_n,v_n]^\top,1\leq n\leq N_t$. The antenna position of the $k^{th}$ user is $\mathbf{p}_k=[x_k,y_k]^\top,1\leq k \leq K$. The positions of LUs' FAs can be flexibly adjusted within a rectangular region $\mathcal{C}_k$ of size $X_{\max} \times Y_{\max}$. According to \cite{MA_channel_model}, the transmit and receive field response vectors (FRV) for the $k^{th}$ LU are represented as 
\begin{equation}
\mathbf{g}_k(\mathbf{b}_n) = \begin{bmatrix} \rho^t_{k,1,n}, \cdots, \rho^t_{k,L_k,n} \end{bmatrix}^\top\in \mathbb{C}^{L_k\times 1}, k\in  \mathcal{K},
\end{equation}
\begin{equation}
\mathbf{f}_k(\mathbf{p}_k) = \begin{bmatrix} \rho^r_{k,1}, \cdots, \rho^r_{k,L_k} \end{bmatrix}^\top\in \mathbb{C}^{L_k\times 1},k\in \mathcal{K},
\end{equation}
where the transmit and receive phase differences across the $k^{th}$ user's $L_k$ paths are given by
\begin{equation}
    \rho^t_{k,l,n}=e^{j\frac{2\pi }{\lambda} u_n\sin(\theta_{k,l})\cos(\phi_{k,l})+v_n\cos\theta_{k,l}},
\end{equation} 
\begin{equation}
    \rho^r_{k,l}=e^{j\frac{2\pi }{\lambda} x_k\sin(\theta_{k,l})\cos(\phi_{k,l})+y_k\cos{\theta_{k,l}}},
\end{equation}
with $\theta_{k,l}$ and $\phi_{k,l}$ representing the azimuth and elevation angle-of-departure (AoD) of the $l^{th}$ path for the $k^{th}$ user. Furthermore, the path response matrix (PRM) between the reference point $\mathbf{u}_0=[0,0]^\top$ of BS and the reference point $\mathbf{b}_{k,0}=[0,0]^\top$ of the $k^{th}$ user is denoted as $\mathbf{\Sigma}_k\in\mathbb{C}^{L_k\times L_k}$. Thus, the channel from the BS to the $k^{th}$ user is expressed by
\begin{equation}
    \mathbf{h}_k=\mathbf{G}_k^H\mathbf{\Sigma}_k\mathbf{f}_k(\mathbf{p}_k),
\end{equation}
where the field response matrix $\mathbf{G}_k=[\mathbf{g}_k(\mathbf{b}_1),\dots,\mathbf{g}_k(\mathbf{b}_{N_T})]$. Additionally, we adopt the line-of-sight (LoS) channel model for the sensing channel for Eve, which can be obtained as
\begin{equation}
\mathbf{h}_e = 
\sqrt{\eta_e}
\begin{bmatrix}
e^{j \frac{2 \pi}{\lambda} \rho^t_l(t_1, \vartheta_e, \varphi_e)}, \dots, e^{j \frac{2 \pi}{\lambda} \rho^t_l(t_{N_t}, \vartheta_e, \varphi_e)}
\end{bmatrix}^\top,
\end{equation}
where $\vartheta_e$ and $\varphi_e$ represent the azimuth and elevation angles between Eve and the BS, and $\eta_e$ denotes the large-scale fading coefficient for Eve.
\subsection{Signal Model}
Assuming that the system uses a 1-layer RSMA scheme, each user's message $m_k$ is divided into two parts, i.e., the common part $m_{k,c}$ and the private part $m_{k,p}$. The private messages are encoded into $\{s_1, s_2,\cdots,s_K\}$, respectively. The common messages are combined and encoded into a common stream $s_{K+1}$. It is assumed that these streams satisfy $\mathbb{E}[s_is_i^H]=1$, and $\mathbb{E}[s_is_j^H]=0, 1\leq i < j \leq K+1$. The transmit beamforming vector for the $k^{th}$ user is denoted as $\mathbf{w}_k$, while the beamforming vector for the common stream is $\mathbf{w}_{K+1}$. 
Consequently, the received signal at the $k^{th}$ user is given by
\begin{equation}
    y_k = \mathbf{h}_k^H\mathbf{w}_ks_k + \sum\limits_{j=1, j\neq k}^{K+1}\mathbf{h}_k^H \mathbf{w}_j s_j + n_k, 
\end{equation}
where $n_k \sim \mathcal{C}\mathcal{N}(0,\sigma_k^2)$ represents the receive noise. In the 1-layer RSMA system, users first decode the common stream while treating the other streams as interference. After decoding the common stream and subtracting it from the received signal, the users proceed to decode their private streams. Therefore, the common rate for the $k^{\text{th}}$ user can be expressed as 
\begin{equation}
r_{c,k} = \log_2\bigg (1+\frac{|\mathbf{h}_k^H\mathbf{w}_c|^2}{\sum\limits_{j=1}^{K} |\mathbf{h}_k^H\mathbf{w}_j|^2 + \sigma_k^2}\bigg).
\end{equation}
To ensure that the common message is decodable for the $k^{th}$ LU, the common message achievable rate $r_{c,k}$ should be higher than the actual transmission rate of the common message, i.e., $r_{c,k}\geq R_c$. Then, by subtracting the common stream, the private stream rate is given by
\begin{equation}
    r_{p,k} = \log_2\bigg (1+\frac{|\mathbf{h}_k^H\mathbf{w}_k|^2}{\sum\limits_{j=1, j \neq k}^{K} |\mathbf{h}_k^H\mathbf{w}_j|^2 + \sigma_k^2}\bigg).
\end{equation}
In contrast, Eve attempts to decode the common message in the first stage and then cancel it. To this end, we should ensure Eve's common eavesdropping capability $e_{c}$ is lower than the actual common rate, i.e., $e_c< R_c$. Thus, Eve can not decode the common stream, which means the undecodable common stream can be regarded as an AN to protect the messages. Accordingly, Eve's intercepted achievable common rate and private rate for the $k^{th}$ LU are given by
\begin{equation}
    e_{c}=\log_2\bigg(1+\frac{|\mathbf{h}_{e}^H \mathbf{w}_c|^2}{\sum\limits^K_{j=1}|\mathbf{h}_{e}^H \mathbf{w}_j|^2+\sigma_e^2}\bigg),
\end{equation}
\begin{equation}
    e_{p,k}=\log_2\bigg(1+\frac{|\mathbf{h}_{e}^H \mathbf{w}_k|^2}{\sum\limits^K_{i=1,i\neq k}|\mathbf{h}_{e}^H \mathbf{w}_i|^2+|\mathbf{h}_{e}^H \mathbf{w}_c|^2+\sigma_e^2}\bigg).
\end{equation}
Therefore, we can derive the secure rate of the whole system as follows.
\begin{equation}
    \mathcal{R_{\text{secure}}} = \sum\limits_{k=1}^{K} [r_{p,k}-e_{p,k}]^+ +\sum\limits_{k=1}^{K}\alpha_k[r_{c}-e_{c}]^+,
\end{equation}
where $\alpha_k$ is the weighted coefficient of common messages for the $k^{th}$ user and $\sum\limits_{k=1}^{K}\alpha_k=1$.

Additionally, for the radar sensing, we adopt the energy radiated toward the target as a sensing metric, which is given by
\begin{equation}
    S(\vartheta_e,\varphi_e)=\sum\limits_{k=1}^{K+1}|\mathbf{h}_e^H \mathbf{w}_k|^2.
\end{equation}

\subsection{Problem Formulation}
This work focuses on beamforming design and positions optimization of LUs' FAs to maximize the secrecy sum rate in the RSMA-ISAC system. Two cases are considered: perfect and imperfect CSI for both the communication and sensing channels. The optimization problem is then formulated as follows.
\subsubsection{Perfect CSI Scenario}
\begin{subequations}
\begin{align}
\max_{ \{\mathbf{w}_k\}_{k=1}^{K+1}, \{\mathbf{p}_k\}_{k=1}^K}&\mathcal{R_{\text{secure}}} \label{p: p1} \\
\quad \text{s.t.} \quad \sum\limits_{k=1}^{K+1}|\mathbf{w}_k|^2 \leq &P_0,  \label{eq: F}  \\
 \quad r_{c,k}\geq R_c,\forall k\in&\mathcal{K},\label{perfect c2} \\ 
 \quad e_{c}\leq R_c,& \label{c3}\\
S(\vartheta_e,\varphi_e) \geq S_0,&\label{c4}\\
 \quad \mathbf{p}_k \in\mathcal{C}_k, \forall k \in& \mathcal{K}\label{c5}, 
\end{align}
\end{subequations}
where $P_0$ is the maximum transmit power and $S_0$ is the predefined beam gain threshold for sensing. Due to (\ref{perfect c2}) and (\ref{c3}), the common stream can serve a dual purpose: it not only interferes with Eve but also provides valid data to the LUs. (\ref{c5}) ensures that the FAs move within the predefined regions.
\subsubsection{Imperfect CSI Scenario}
In this case, a general scenario is considered, where the BS is assumed to have imperfect CSI for both the communication and sensing channels. The deterministic CSI error model is expressed as
\begin{equation}
    \mathbf{h}_k=\hat{\mathbf{h}}_k+\Delta \mathbf{h}_k,
\end{equation}
where $\hat{\mathbf{h}}_k$ is the estimated channel and the channel error satisfies $\Delta \mathbf{h}_k^\top \Delta \mathbf{h}_k \leq \epsilon_k^2$ \cite{fu2020robust}. Additionally, in terms of the imperfect sensing channel, the angles with estimation errors for Eve are represented by the two sets $\Pi_{\vartheta_e}$ and $\Pi_{\varphi_e}$, where $N_{e,1}$ and $N_{e,2}$ denote the element number of each set. Considering the imperfect CSI of both communication and sensing channels, we formulate a robust optimization problem to maximize the worst-case secure rate as follows
\begin{subequations}
\begin{align}
\max_{ \{\mathbf{w}_k\}_{k=1}^{K+1}, \{\mathbf{p}_k\}_{k=1}^K}&\min_{\{\Delta \mathbf{h}_k\}_{k=1}^K,\vartheta_e,\varphi_e}\mathcal{R_{\text{secure}}} \label{p: p2} \\
\quad \text{s.t.}  
\quad  \quad e_c &\leq R_c, \forall\vartheta_e\in \Pi_{\vartheta_e},\forall\varphi_e\in\Pi_{\varphi_e},\label{imperfect c3}\\
S(\vartheta_e,\varphi_e)&\geq S_0,\forall\vartheta_e\in \Pi_{\vartheta_e},\forall\varphi_e\in\Pi_{\varphi_e},\label{imperfect c4}\\
&(\ref{eq: F}), (\ref{perfect c2}), (\ref{c5}).\nonumber
\end{align}
\end{subequations}

%% file: Chapter/proposed-solution.tex
\section{Proposed Design Under Perfect CSI}
\label{perfect CSI}

Directly solving (\ref{p: p1}) is challenging due to the non-convexity of the optimization function $\mathcal{R}_{\text{secure}}$ with respect to $\{\mathbf{w}_k\}_{k=1}^{K+1}$ and $\{\mathbf{p}_k\}_{k=1}^K$. To address this, we propose an AO algorithm that optimizes the beamforming vectors and FA positions separately through a series of transformations. Specifically, the problem is first reformulated using SDP, followed by the application of SCA to optimize the surrogate upper bound. The beamforming vectors and FA positions are alternately optimized in the subproblems. Finally, the overall computational complexity of the proposed algorithm is analyzed.

\subsection{SDP Reformulation}
To enhance the tractability of the optimization problem, we employ SDP by introducing a rank-one constraint, which transforms the beamforming vectors into rank-one matrices. The reformulated problem is expressed as follows. 
\begin{subequations}
\begin{align}
\max_{ \{\mathbf{W}_i\}_{i=1}^{K+1},\{\mathbf{p}_k\}_{k=1}^K}\sum_{k=1}^K&\mathcal{R_{\text{secure$,k$}}}(\{\mathbf{W}_i\}_{i=1}^{K+1},\{\mathbf{p}_k\}_{k=1}^K) \label{p: p3}\\
\quad \text{s.t.}  \sum_{i=1}^{K+1}&\text{Tr}(\mathbf{W}_i) \leq P_0,  \label{power constraint}  \\
\quad   
-\sum\limits_{i=1}^{K+1}\mathbf{h}_k^H\mathbf{W}_i\mathbf{h}_k-\sigma_k^2+&2^{R_c}\big(\sum\limits_{i=1}^K\mathbf{h}_k^H\mathbf{W}_i\mathbf{h}_k+\sigma_k^2\big)\leq 0,\nonumber\\
&\quad\quad\quad\quad\quad\quad\quad\forall k\in\mathcal{K}, \label{cons for ru}\\ 
\sum\limits_{i=1}^{K+1}\mathbf{h}_e^H\mathbf{W}_i\mathbf{h}_e+\sigma_e^2-&2^{R_c}\big(\sum\limits_{i=1}^K\mathbf{h}_e^H\mathbf{W}_i\mathbf{h}_e+\sigma_e^2\big)\leq 0,\nonumber\\
&\quad\quad\quad\quad\quad\quad\quad\forall k\in\mathcal{K},\label{cons for re}\\
\sum\limits_{i=1}^{K+1}\mathbf{h}_e^H&\mathbf{W}_i\mathbf{h}_e\geq S_0,\label{cons for G0}\\
\text{Rank}(\mathbf{W}_i)=1&,1\leq i\leq K+1, \label{rank1 cons}\\
\quad & (\ref{c5}). \nonumber
\end{align}
\end{subequations}
where $\mathcal{R_{\text{secure$,k$}}}$ is shown in (\ref{perfect original objective}). 
\begin{figure*}[t]
    \begin{align}  R_{\text{Secure$,k$}}&=\alpha_k\bigg[R_c+\log_2\bigg(\sum\limits_{i=1}^K\mathbf{h}_e^H\mathbf{W}_i\mathbf{h}_e+\sigma_e^2\bigg)\bigg]
        +\log_2\bigg(\sum\limits_{i=1}^K\mathbf{h}_k^H\mathbf{W}_i\mathbf{h}_k+\sigma_k^2\bigg)
        +\log_2\bigg(\sum\limits_{i=1,i\neq k}^{K+1}\mathbf{h}_e^H\mathbf{W}_i\mathbf{h}_e+\sigma_e^2\bigg) 
        \nonumber\\
        &\underbrace{-\log_2\bigg(\sum\limits_{i=1,i\neq k}^K\mathbf{h}_k^H\mathbf{W}_i\mathbf{h}_k+\sigma_k^2\bigg)}_{\Phi_{k,1}}
        \underbrace{-(\alpha_k+1)\log_2\bigg(\sum\limits_{i=1}^{K+1}\mathbf{h}_e^H\mathbf{W}_i\mathbf{h}_e+\sigma_e^2\bigg)}_{\Phi_{k,2}},
        \label{perfect original objective}
    \end{align}
    
\end{figure*}

\subsection{Dual-functional Beamforming Design}
We first fix the FA positions and then optimize the beamforming matrices. The dual-functional beamforming solution is obtained by solving the following problem.
\begin{subequations}
\begin{align}
\max_{ \{\mathbf{W}_i\}_{i=1}^{K+1}}\sum_{k=1}^K\mathcal{R_{\text{secure$,k$}}}&(\{\mathbf{W}_i\}_{i=1}^{K+1},\{\mathbf{p}_k\}_{k=1}^K)\\
\quad \text{s.t.} \quad (\ref{power constraint})-&(\ref{rank1 cons}). \nonumber
\end{align}
\end{subequations}
To further address this problem, we employ the linearization technique in \cite{lineartechnique} to iteratively approximate and optimize the objective function.

\noindent \textbf{Lemma 1} : \textit{Assume a positive scalar \( m \) and \( f(m) = -\frac{mn}{\ln 2} + \log_2 m + \frac{1}{\ln 2} \). We have}
\begin{equation}
    - \log_2 n = \max_{m > 0} f(m), \label{proposition1}
\end{equation}
\textit{and the optimal value of \( m \) that maximizes the function on the right-hand side of equation (\ref{proposition1}) is given by \( m = \frac{1}{n} \). This lemma can be proved by solving for the value of \( m \) that makes the derivative of \( f(m) \) equal to zero.}

According to \textbf{Lemma 1}, $\Phi_{k,1}$ and $\Phi_{k,2}$ can be approximate using the surrogate lower bound as follows.
\begin{equation}
    \Phi_{k,1}\geq -\frac{m_{k}}{\ln 2}(\sum\limits_{i=1,i\neq k}^K\mathbf{h}_k^H\mathbf{W}_i\mathbf{h}_k+\sigma_k^2)
    +\log_2 m_{k}+\frac{1}{\ln 2}, \label{m equation}
\end{equation}
\begin{equation}
    \Phi_{k,2}\geq  (\alpha_k+1)[-\frac{l_{k}}{\ln 2}(\sum\limits_{i=1}^{K+1}\mathbf{h}_e^H\mathbf{W}_i\mathbf{h}_e+\sigma_e^2)+\log_2 l+\frac{1}{\ln 2}],\label{l equation}
\end{equation}
where the optimal auxiliary variables $m_k$ and $l$, are given as
\begin{equation}
    m_k^*=(\sum\limits_{i=1,i\neq k}^K\mathbf{h}_k^H\mathbf{W}_i\mathbf{h}_k+\sigma_k^2)^{-1},\forall k\in\mathcal{K},\label{update m}
\end{equation}
\begin{equation}
    l^*=(\sum\limits_{i=1}^{K+1}\mathbf{h}_e^H\mathbf{W}_i\mathbf{h}_e+\sigma_e^2)^{-1}.\label{update l}
\end{equation}
Thus, the objective function can be transformed into convex form (\ref{perfect transformed objective}). It is worth noting that $\{\mathbf{W}_i\}_{i=1}^{K+1}$ can be proved to satisfy the rank-one property by \cite[Theorem 1]{fu2020robust}. Consequently, by dropping the constraint (\ref{rank1 cons}), the reformulated problem becomes

\begin{figure*}[t]
    
    \begin{align}
    &\hat{R}_{\text{Secure$,k$}}=\underbrace{\alpha_k\bigg[R_c+\log_2\bigg(\sum\limits_{i=1}^K\mathbf{h}_e^H\mathbf{W}_i\mathbf{h}_e
    +\sigma_e^2\bigg)\bigg]}_{\Psi_{k,1}}
    +\underbrace{\log_2\left( \sum\limits_{i=1}^K \mathbf{h}_k^H \mathbf{W}_i \mathbf{h}_k + \sigma_k^2 \right)}_{\Xi_{k,1}}
    +\underbrace{\log_2\bigg(\sum\limits_{i=1,i\neq k}^{K+1}\mathbf{h}_e^H\mathbf{W}_i\mathbf{h}_e
    +\sigma_e^2\bigg)}_{\Psi_{k,2}} \nonumber\\ 
    &-\frac{1}{ln(2)}m_k( \underbrace{\sum\limits_{i=1,i\neq k}^K \mathbf{h}_k^H \mathbf{W}_i \mathbf{h}_k + \sigma_k^2}_{\Xi_{k,2}})+\underbrace{\log_2m_k+\frac{1}{ln(2)}
    +(\alpha_k+1)\bigg[-\frac{1}{ln(2)}l(\sum\limits_{i=1}^{K+1}\mathbf{h}_e^H\mathbf{W}_i\mathbf{h}_e+\sigma_e^2)+\log_2l+\frac{1}{ln(2)}\bigg]}_{\Psi_{k,3}}.
    \label{perfect transformed objective}
\end{align} 

\hrulefill
\end{figure*}

\begin{subequations}
\begin{align}
\max_{ \{\mathbf{W}_i\}_{i=1}^{K+1}}\sum_{k=1}^K\mathcal{\hat{R}_{\text{secure$,k$}}}&(\{\mathbf{W}_i\}_{i=1}^{K+1},\{\mathbf{p}_k\}_{k=1}^K) \label{p: perfect beam}\\
\quad \text{s.t.} \quad (\ref{power constraint})-&(\ref{cons for G0}).\nonumber
\end{align}
\end{subequations}
The resulting problem is convex and can be efficiently solved using the CVX optimization toolbox. Subsequently, the auxiliary variables are updated according to (\ref{update m}) and (\ref{update l}). The beamforming matrix and auxiliary variables are alternately optimized until convergence is achieved.

\subsection{Antenna Position Optimization}
For any given beamforming matrices, the optimization problem for antenna positions is given as follows. Each of these subproblems must be addressed sequentially.

\begin{subequations}
\begin{align}
\max_{ \{\mathbf{p}_k\}_{k=1}^K}\sum_{k=1}^K&\hat{\mathcal{R}}_{\text{secure$,k$}}(\{\mathbf{W}_i\}_{i=1}^{K+1},\{\mathbf{p}_k\}_{k=1}^K) \\
\quad \text{s.t.}  & \quad (\ref{cons for ru}), (\ref{c5}).  \nonumber
\end{align}
\end{subequations}
The objective function and constraint (\ref{cons for ru}) are non-convex with respect to the transmit power variables \(\{\mathbf{p}_i\}_{k=1}^K\). To address this challenge, we handle these non-convex components subsequently.

\subsubsection{Convex Reformulation of Objective Function}
In the transformed objective function in (\ref{perfect transformed objective}), $\Xi_{k,1}$ and $\Xi_{k,2}$ are neither convex nor concave w.r.t. $\{\mathbf{p}_i\}_{k=1}^{K}$. To address this, we employ the SCA method. Let the following constraint hold
\begin{equation}
\sum\limits_{i=1}^{K+1}\mathbf{h}_k^H\mathbf{W}_i\mathbf{h}_k+\sigma_k^2\geq 2^{\tau_{k,1}}, \label{tau1}
\end{equation}
Expanding this constraint yields
\begin{equation}
\mathbf{f}_k^H(\mathbf{p}_k)\mathbf{E}_{k,1}\mathbf{f}_k(\mathbf{p}_k)+2^{\tau_{k,1}}-\sigma_k^2\leq 0,\label{fe equation}
\end{equation}
where $\mathbf{E}_{k,1}=-\mathbf{\Sigma}_k^H\mathbf{G}_k^H\mathbf{\Sigma}_k$. Since this constraint remains non-convex, we utilize a lemma introduced in \cite{ma2024movable} to reformulate (\ref{fe equation}).

\noindent \textbf{Lemma 2} : \textit{For given \( \mathbf{p}_k^{(t)} \) in the \( t \)-th iteration, we have}
\begin{align}
&\mathbf{f}_k^\top(\mathbf{p}_k) \mathbf{E}_k \mathbf{f}_k(\mathbf{p}_k) \leq 
\mathbf{f}_k^\top(\mathbf{p}_k) \mathbf{\Lambda}_k \mathbf{f}_k(\mathbf{p}_k) 
\nonumber\\
&- 2\Re \left\{ \mathbf{f}_k^\top(\mathbf{p}_k)(\mathbf{\Lambda}_k - \mathbf{E}_k) \mathbf{f}_k(\mathbf{p}_k^{(t)}) \right\} 
\nonumber\\
&+ \mathbf{f}_k^\top(\mathbf{p}_k^{(t)}) (\mathbf{\Lambda}_k - \mathbf{E}_k) \mathbf{f}_k(\mathbf{p}_k^{(t)}) ,
\end{align}
\textit{where} \( \mathbf{\Lambda}_k = \lambda_{\max}(\mathbf{E}_k)\mathbf{I} \).

\textit{Proof: Refer to \cite[Lemma 1]{ma2024movable}.}

According to \textbf{Lemma 2}, the surrogate upper bound for the left-hand function of (\ref{fe equation}) is given by
\begin{align}
&J(\mathbf{p}_k)\triangleq
\mathbf{f}_k^\top(\mathbf{p}_k) \mathbf{\Lambda}_{k,1} \mathbf{f}_k(\mathbf{p}_k) 
\nonumber\\
&- 2\Re \left\{ \mathbf{f}_k^\top(\mathbf{p}_k)(\mathbf{\Lambda}_{k,1} - \mathbf{E}_{k,1}) \mathbf{f}_k(\mathbf{p}_k^{(t)}) \right\} 
\nonumber\\
&+ \mathbf{f}_k^\top(\mathbf{p}_k^{(t)}) (\mathbf{\Lambda}_{k,1} - \mathbf{E}_{k,1}) \mathbf{f}_k(\mathbf{p}_k^{(t)})+2^{\tau_{k,1}}-\sigma_k^2 ,
\end{align}
where $\mathbf{\Lambda}_{k,1}=\lambda_{\max}(\mathbf{E}_{k,1})\mathbf{I}$. Since $\mathbf{f}_k^\top(\mathbf{p}_k) \mathbf{\Lambda}_{k,1} \mathbf{f}_k(\mathbf{p}_k)=\lambda_{\max}(\mathbf{E}_{k,1})L_k$ is constant, we can simplify $J(\mathbf{p}_k)$ as
\begin{align}
J(\mathbf{p}_k|\mathbf{p}_k^{(t)})&=\lambda_{\max}(\mathbf{E}_{k,1})L_k- 2\Re \left\{ \mathbf{f}_k^\top(\mathbf{p}_k)\mathbf{q}_{k,1} \right\} \nonumber\\
&+ \mathbf{f}_k^\top(\mathbf{p}_k^{(t)}) \mathbf{q}_{k,1}+2^{\tau_{k,1}}-\sigma_k^2,
\end{align}
where $\mathbf{q}_{k,1}=(\mathbf{\Lambda}_{k,1} - \mathbf{E}_{k,1}) \mathbf{f}_k(\mathbf{p}_k^{(t)})$.
The second-order Taylor expansion is employed to obtain the surrogate function. By introducing $\delta_{k,1}\mathbf{I}_2\succeq \nabla^2 J(\mathbf{p}_k)$, we approximate $J(\mathbf{p}_k)$ at $\mathbf{p}_k$ as follows.
\begin{align}
    J(\mathbf{p}_k|\mathbf{p}_k^{(t)})&\leq \hat{J}(\mathbf{p}_k|\mathbf{p}_k^{(t)})\triangleq J(\mathbf{p}^{(t)}_k)+\nabla J(\mathbf{p}^{(t)}_k)^\top(\mathbf{p}_k-\mathbf{p}_k^{(t)})\nonumber\\
    &+\frac{\delta_{k,1}}{2}(\mathbf{p}_k-\mathbf{p}_k^{(t)})^\top(\mathbf{p}_k-\mathbf{p}_k^{(t)}),
    \label{J upper}
\end{align}
where $\nabla J(\mathbf{p}_k)$ and $\delta_{k,1}$ are provided in Appendix \ref{Appendix perfect antenna 1}.

Next, we focus on the term $\Xi_{k,2}$. Define
\begin{equation}
\mathbf{E}_{k,2}=\mathbf{\Sigma}_k^H\mathbf{G}_k(\sum\limits_{i\neq k}^{K}\mathbf{W}_i)\mathbf{G}_k^H\mathbf{\Sigma}_k,
\end{equation}
\begin{equation}
    \mathbf{\Lambda}_{k,2}=\lambda_{\max}(\mathbf{E}_{k,2})\mathbf{I}.
\end{equation}
Following \textbf{Lemma 2}, we derive the surrogate upper bound for $\Xi_{k,2}$ as
\begin{align} 
    \Xi_{k,2}\leq T_k(\mathbf{p}_k) \triangleq \lambda_{\max}(\mathbf{E}_{k,2})L_k
    &-2\Re\{\mathbf{f}_k^H(\mathbf{p}_k)\mathbf{q}_{k,2}\} \nonumber\\
    &+\mathbf{f}_k^H(\mathbf{p}^{(t)}_k)\mathbf{q}_{k,2}+\sigma_k^2,
\end{align}
where $\mathbf{q}_{k,2}=(\mathbf{\Lambda}_{k,2}-\mathbf{E}_{k,2})\mathbf{f}_k^H(\mathbf{p}^{(t)}_k)$.
Similarly, by introducing $\delta_{k,2}\succeq \nabla^2 T_k(\mathbf{p}_k)$ and applying the second-order Taylor expansion theorem, we obtain the following upper bound
\begin{align}
    T_k(\mathbf{p}_k) &\leq \hat{T}_k(\mathbf{p}_k|\mathbf{p}_k^{(t)})\triangleq T_k(\mathbf{p}^{(t)}_k)+\nabla T_k(\mathbf{p}^{(t)}_k)(\mathbf{p}_k-\mathbf{p}_k^{(t)})\nonumber\\
    &+\delta_{k,2}(\mathbf{p}_k-\mathbf{p}_k^{(t)})^\top(\mathbf{p}_k-\mathbf{p}_k^{(t)}),
    \label{T upper}
\end{align}
where $\nabla T_k(\mathbf{p}_k)$ and $\delta_{k,2}$ are provided in Appendix \ref{Appendix T upper}.

\subsubsection{Convex Reformulation of Constraint (\ref{cons for ru})}
We extend the application of $\textbf{Lemma 2}$ to the left-hand side function of constraint (\ref{cons for ru}).
\begin{align}
     D_k(\mathbf{p}_k)&\triangleq \lambda_{\max}(\mathbf{R}_k)L_k-2\Re\{\mathbf{f}_k^H(\mathbf{p}_k)(\mathbf{\Omega}_k-\mathbf{R}_k)\mathbf{f}_k(\mathbf{p}_k^{(t)})\}\nonumber\\
    &+\mathbf{f}_k^H(\mathbf{p}^{(t)}_k)(\mathbf{\Omega}_k-\mathbf{R}_k)\mathbf{f}_k(\mathbf{p}_k^{(t)})+(2^{R_c}-1)\sigma_k^2\nonumber\\
    &=\lambda_{\max}(\mathbf{R}_k)L_k+(2^{R_c}-1)\sigma_k^2+2\Re\{\mathbf{f}_k^H(\mathbf{p}_k)\boldsymbol{a}_k\}\nonumber\\
    &+\mathbf{f}_k^H(\mathbf{p}^{(t)}_k)\boldsymbol{a}_k,
\end{align}
where
$\mathbf{R}_k=\mathbf{\Sigma}_k^H\mathbf{G}_k[(2^{R_c}-1)\sum\limits_{i=1}^{K}\mathbf{W}_i-\mathbf{W}_{K+1}]\mathbf{G}^H_k\mathbf{\Sigma}_k
$, $\mathbf{\Omega}_k=\lambda_{\max}(\mathbf{R}_k)\mathbf{I}$ and $\boldsymbol{a}_k=(\mathbf{\Omega}_k-\mathbf{R}_k)\mathbf{f}_k(\mathbf{p}_k^{(t)})$. With a similar derivation, we get the surrogate upper bound $\hat{D}_k(\mathbf{p}_k|\mathbf{p}_k^{(t)})$ as 
\begin{align}
    D_k(\mathbf{p}_k)\leq \hat{D}_k(&\mathbf{p}_k|\mathbf{p}_k^{(t)}) \triangleq D_k(\mathbf{p}^{(t)}_k)(\mathbf{p}_k-\mathbf{p}_k^{(t)})\nonumber\\
    &+\frac{\delta_{k,3}}{2}(\mathbf{p}_k-\mathbf{p}_k^{(t)})^\top(\mathbf{p}_k-\mathbf{p}_k^{(t)})\leq 0. 
    \label{D upper}
\end{align}
The gradient $\nabla D_k(\mathbf{p}_k)$ and $\delta_{k,3}$ are shown in Appendix \ref{Appendix D upper}.

By substituting the original objective function and constraint (\ref{cons for ru}) with the upper-bound surrogate functions (\ref{J upper}), (\ref{T upper}), and (\ref{D upper}), we reformulate the problem as the following convex optimization problem
\begin{subequations}
\begin{align}
\max_{ \{\mathbf{p}_k\}_{k=1}^K}\sum_{k=1}^K(\Psi_{k,1}+&\Psi_{k,2}+\Psi_{k,3}+\hat{J}(\mathbf{p}_k|\mathbf{p}_k^{(t)})+\hat{T}_k(\mathbf{p}_k|\mathbf{p}_k^{(t)})) \label{p: perfect antenna}\\
\quad \text{s.t.}  & \quad \hat{D}(\mathbf{p}_k|\mathbf{p}_k^{(t)})) \leq 0,\quad(\ref{c5}).  \nonumber
\end{align}
\end{subequations}
This is a convex problem and can be efficiently solved using standard techniques. Hence, the algorithm for obtaining high-quality solutions of the beamforming matrices and positions of FA is summarized in \textbf{Algorithm \ref{algo1}}.
\begin{algorithm}[t]
    \renewcommand{\algorithmicrequire}{\textbf{Initialization:}}
	\renewcommand{\algorithmicensure}{\textbf{Output:}}
    \caption{AO for Solving Problem (\text{P3}).}
     \label{algo1}
    \begin{algorithmic}[1]
        \REQUIRE Set the convergence tolerances $\zeta_1=\zeta_2=\zeta_3=10^{-4}$, the outer iteration index $N_1=0$, beam updating iteration index $N_2=0$ and FA updating iteration index $N_3=0$. Initialize $\{\mathbf{p}_k^{(0)}\}_{k=1}^K$ and auxiliary parameters $\{m^{(0)}\}_{k=1}^K$, $l^{(0)}$, maximum iteration index $N_1^{\max}$, $N_2^{\max}$ and $N_3^{\max}$.
        \REPEAT
            \REPEAT 
                \STATE Update $\{\mathbf{W}_k^{(N_2+1)}\}_{k=1}^{K+1}$ by solving (\ref{p: perfect beam}).
                \STATE Update $\{m^{(N_2+1)}\}_{k=1}^K$ and $l^{(N_2+1)}$ by (\ref{update m}), (\ref{update l}).
                \STATE Update iteration index $N_1\xleftarrow{}N_1+1$.
            \UNTIL{ the increase of the objective value in (\ref{p: perfect beam}) is  below the threshold $\zeta_1$ or $N_1>N_1^{\max}$}.
            
            \REPEAT 
                \STATE Update $\{\mathbf{p}_k^{(N_3+1)}\}_{k=1}^K$ by solving (\ref{p: perfect antenna}).
                \STATE Update iteration index $N_2\xleftarrow{}N_2+1$.
            \UNTIL{the increase of the objective value in (\ref{p: perfect antenna}) is  below the threshold $\zeta_2$ or $N_2>N_2^{\max}$}. 
             \STATE Update iteration index $N_3\xleftarrow{}N_3+1$, $N_1\xleftarrow{}0, N_2\xleftarrow{}0.$
     \UNTIL{the $|\hat{R}^{(N_3)}_{\text{Secure}}-\hat{R}^{(N_3-1)}_{\text{Secure}}| <= \zeta_3$ or $N_3>N_3^{\max}$.}
        \ENSURE $\hat{R}^{(N_3)}_{\text{Secure}}$ and $\mathbf{W}_k^*=\mathbf{w}^*_k\mathbf{w}^{*H}_k$.
    \end{algorithmic}
\end{algorithm}
\subsection{Overall Solution}
Convergence Analysis: In \textbf{Algorithm \ref{algo1}}, each subproblem is solved locally, ensuring that the objective function (\ref{perfect transformed objective}) is non-decreasing over iterations. Additionally, since (\ref{perfect transformed objective}) is upper-bounded by the finite transmit power budget, the solution generated by \textbf{Algorithm \ref{algo1}} is guaranteed to converge.

Computational Complexity: The computational complexity is mainly dominated by solving the SCA problems in (\ref{p: perfect antenna}) and (\ref{p: perfect beam}). From \cite{fu2020robust}, the computational complexity for solving (\ref{p: perfect antenna}) is $ \mathcal{O}\left( N^{\frac{1}{2}} \left( M N^3 + M^2 N^2 + M^3 N \right) \log \left( \frac{1}{\zeta_1} \right) \right)$, where $M=K+3$ and $N=N_T+1$, and $\zeta_1$ is the accuracy. The complexity of obtaining maximum eigenvalues $\boldsymbol{\Lambda}_{k,1},\boldsymbol{\Lambda}_{k,2}$ and $\boldsymbol{\Omega}$ are $\mathcal{O}(L_k^3)$. The complexities for calculating $\nabla \mathbf{J},\nabla \mathbf{T}_k$ and $ \nabla \mathbf{R}_k$ are $\mathcal{O}(L_k)$.The complexity of solving (\ref{p: perfect antenna}) is $\mathcal{O}\left(N_T^{1.5}\ln(\frac{1}{\zeta_2})\right)$, where $\zeta_2$ is the accuracy. Thus, the overall complexity is $\mathcal{O}\big(N_3N_1[(N_T+1)^{3.5}(K+3)+(K+3)^2(N_T+1)^{2.5}+(K+3)^3(N_T+1)\ln(\frac{1}{\zeta_1})]$  $+N_3N_2[KL_k^3+N_T^{1.5}\ln(\frac{1}{\zeta_2})]\big) $, where $N_1,N_2$ are the inner iteration number and $N_3$ is the outer iteration number.

\section{Proposed Design Under Imperfect CSI}
 We use \( t_k \) as a surrogate lower bound for the original problem (\ref{p: p1}), simplifying the max-min problem in (\ref{p: p2}) by introducing the constraint (\ref{cons tk}). Similar to the solution under perfect conditions, a new problem (\ref{p: p4}) is derived as follows.
\label{imperfect CSI}
\begin{subequations}
\begin{align}
\max_{ \{\mathbf{W}_i\}_{i=1}^{K+1}, \{\mathbf{p}_k\}_{k=1}^K}& \quad \sum\limits_{k=1}^{K}t_k \label{p: p4}\\
 \text{s.t.} \quad \hat{\mathcal{R}}_{\text{secure,$k$}}&(\{\mathbf{W}_i\}_{i=1}^{K+1},\{\mathbf{p}_k\}_{k=1}^K) \geq t_k,\nonumber\\
 \forall k\in\mathcal{K},\forall\{\Delta \mathbf{h}_k&\}_{k=1}^K,\forall\vartheta_e\in \Pi_{\vartheta_e},\forall\varphi_e\in\Pi_{\varphi_e},\label{cons tk}\\
 \quad  (\hat{\mathbf{h}}_k+\Delta \mathbf{h}_k)^H\mathbf{M}(\hat{\mathbf{h}}_k&+\Delta \mathbf{h}_k)+(2^{R_c}-1)\sigma_k^2\leq 0, \label{imperfect ru}\\ 
\sum\limits_{i=1}^{K+1}\mathbf{h}_e^H\mathbf{W}_i\mathbf{h}_e+\sigma_e^2-&2^{R_c}\big(\sum\limits_{i=1}^K\mathbf{h}_e^H\mathbf{W}_i\mathbf{h}_e+\sigma_e^2\big)\leq 0,\nonumber\\ &\forall\vartheta_e\in \Pi_{\vartheta_e},\forall\varphi_e\in\Pi_{\varphi_e}, 
\label{cons for imperfect re}\\
\sum\limits_{i=1}^{K+1}\mathbf{h}_e^H\mathbf{W}_i\mathbf{h}_e\geq S_0,&\forall\vartheta_e\in \Pi_{\vartheta_e},\forall\varphi_e\in\Pi_{\varphi_e}, \label{cons for imperfect sensing}\\
(\ref{power constraint}), &(\ref{rank1 cons}), (\ref{c5}),\nonumber
\end{align}
\end{subequations}
where $\mathbf{M}=(2^{R_c}-1)\sum\limits_{i=1}^K \mathbf{W}_i-\mathbf{W}_{K+1}$. Compared to problem (\ref{p: p3}), the formulated problem (\ref{p: p4}) under imperfect CSI conditions introduces significantly greater analytical complexity.

\subsection{Dual-functional Beamforming Design}
This complexity arises from its max-min problem structure, coupled with an infinite number of inequality constraints due to the continuous channel error $\{\Delta \mathbf{h}\}_{k=1}^K$. Consequently, the solution methodology developed for perfect CSI conditions cannot be directly applied. To address this difficulty, we leverage \textbf{Lemma 3 (S-procedure)} to reformulate the originally intractable problem into a more manageable form with a finite number of constraints.

\textbf{Lemma 3 (S-Procedure)}: \textit{Let a function \( f_i(\mathbf{x}),\ i \in \{1, 2\},\ \mathbf{x} \in \mathbb{C}^{N \times 1} \), which is defined as}
\begin{equation}
    f_i(\mathbf{x}) = \mathbf{x}^H \mathbf{A}_i \mathbf{x} + 2 \Re \left\{ \mathbf{b}_i^H \mathbf{x} \right\} + c_i,
\end{equation}
\textit{where \( \mathbf{A}_i \in \mathbb{C}^{N\times N} \), \( \mathbf{b}_i \in \mathbb{C}^{N \times 1} \), and \( c_i \in \mathbb{R} \). Then, the implication \( f_1(\mathbf{x}) \leq 0 \Rightarrow f_2(\mathbf{x}) \leq 0 \) holds if and only if there exists a \( \delta \geq 0 \) such that}
\begin{equation}
    \delta \begin{bmatrix}
        \mathbf{A}_1 & \mathbf{b}_1 \\
        \mathbf{b}_1^H & c_1
    \end{bmatrix} - \begin{bmatrix}
        \mathbf{A}_2 & \mathbf{b}_2 \\
        \mathbf{b}_2^H & c_2
    \end{bmatrix} \succeq \mathbf{0}.
\end{equation}

\textit{Proof: Refer to \cite{boyd2004convex}.}

\subsubsection{Convex Reformulation of Constraint (\ref{cons tk})}
Due to the continuity of \( \Delta \mathbf{h}_k \), the condition (\ref{cons tk}) involves an infinite number of inequalities.  To address this, we focus on \( \Xi_{k,1} \) and \( \Xi_{k,2} \) in \( \hat{R}_{\text{Secure},k} \). First, we expand (\ref{tau1}) in this imperfect case, which yields 
\begin{align}
    -\Delta \mathbf{h}_k^H \sum\limits_{i=1}^{K} \mathbf{W}_i \Delta &\mathbf{h}_k 
    -2\Re\left\{\hat{\mathbf{h}_k}^H \sum\limits_{i=1}^{K} \mathbf{W}_i \Delta \mathbf{h}_k\right\} \nonumber\\
    &-\hat{\mathbf{h}_k}^H \sum\limits_{i=1}^{K} \mathbf{W}_i \hat{\mathbf{h}_k}
    +2^{\tau_{k,3}} - \sigma_k^2 \leq 0. \label{expanded tau1}
\end{align}
Thus, leveraging the S-procedure, we assume the existence of a scalar $o_{k,1} \geq 0$, such that (\ref{expanded tau1}) is equivalent to the following linear matrix inequality (LMI)
\begin{equation}
\scalebox{0.85}{$
    o_{k,1} \left[ \begin{matrix}
\mathbf{I} & \mathbf{0} \\
\mathbf{0} & -\epsilon_k^2
\end{matrix} \right]-\left[ \begin{matrix}
-\sum\limits_{i=1}^K \mathbf{W}_i & -\hat{\mathbf{h}}_k^H\sum\limits_{i=1}^K \mathbf{W}_i \\
-\sum\limits_{i=1}^K \mathbf{W}_i^H\hat{\mathbf{h}}_k & -\hat{\mathbf{h}}_k^H\sum\limits_{i=1}^K \mathbf{W}_i\hat{\mathbf{h}}_k+\tau_{k,3}-\sigma_k^2
\end{matrix} \right] \succeq \mathbf{0}. \label{cons for S o1}
$}
\end{equation}
Similarly, in terms of $\Xi_{k,2}$, we assume the following constraint
\begin{equation}
\mathbf{h}_k^H\sum\limits_{i=1,i\neq k}^{K}\mathbf{W}_i\mathbf{h}_k+\sigma_k^2\leq \tau_{k,4}. \label{tau2 cons}
\end{equation}
Applying the S-procedure again, we assume the existence of a scalar $o_{k,2} \geq 0$, such that (\ref{tau2 cons}) is equivalent to

\begin{equation}
\scalebox{0.83}{$
    o_{k,2} \left[ \begin{matrix}
\mathbf{I} & \mathbf{0} \\
\mathbf{0} & -\epsilon_k^2
\end{matrix} \right]
-\left[ \begin{matrix}
\sum\limits_{i=1,i\neq k}^{K} \mathbf{W}_i & \hat{\mathbf{h}}_k^H \sum\limits_{i=1,i\neq k}^{K} \mathbf{W}_i \\
\sum\limits_{i=1,i\neq k}^{K} \mathbf{W}_i^H \hat{\mathbf{h}}_k & \hat{\mathbf{h}}_k^H \sum\limits_{i=1,i\neq k}^{K} \mathbf{W}_i \hat{\mathbf{h}}_k+\sigma_k^2-\tau_{k,4}
\end{matrix} \right] \succeq \mathbf{0}. \label{cons for S o2}
$}
\end{equation}

Then, we assume the lower bounds of $\Psi_1$ and $\Psi_2$ for different $\vartheta_e$ and $\varphi_e$ are $\hat{\Psi}_{k,1}$ and $\hat{\Psi}_{k,2}$, respectively, which means
\begin{equation}
\hat{\Psi}_{k,1}\leq \Psi_1, \forall k\in\mathcal{K},\forall\vartheta_e\in \Pi_{\vartheta_e},\forall\varphi_e\in\Pi_{\varphi_e}, \label{cons:Psi1lowerbound}
\end{equation}
\begin{equation}
\hat{\Psi}_{k,2}\leq \Psi_2, \forall k\in\mathcal{K},  \forall\vartheta_e\in \Pi_{\vartheta_e},\forall\varphi_e\in\Pi_{\varphi_e}.\label{cons:Psi2lowerbound}
\end{equation}
It is assumed that 
\begin{equation}
\sum\limits_{i=1}^{K+1}\mathbf{h}_e^H\mathbf{W}_i\mathbf{h}_e+\sigma_e^2\leq \beta, \forall\vartheta_e\in \Pi_{\vartheta_e},\forall\varphi_e\in\Pi_{\varphi_e}.\label{cons:Psi3lowerbound}
\end{equation}
Then we can obtain the lower bound for $\Psi_3$, i.e., $\Psi_3\geq\hat{\Psi}_{k,3}=\log_2m_k+\frac{1}{ln(2)}
+(\alpha_k+1)\bigg[-\frac{1}{ln(2)}l\beta+\log_2l+\frac{1}{ln(2)}\bigg], \forall k\in\mathcal{K}, \forall\vartheta_e\in \Pi_{\vartheta_e},\forall\varphi_e\in\Pi_{\varphi_e}$.

\subsubsection{Convex Reformulation of Constraint (\ref{imperfect ru})}
It can be expanded as 
\begin{equation}
    \Delta \mathbf{h}_k^H\mathbf{M}\Delta \mathbf{h}_k +2\Re\{\hat{\mathbf{h}}^H\mathbf{M}\Delta \mathbf{h}_k\}+\mathbf{\hat{h}}_k^H\mathbf{M}\mathbf{\hat{h}}_k+(2^{R_c}-1)\sigma_k^2\leq 0, \label{cons for imperfect ru}
\end{equation}
Applying \textbf{Lemma 3}, there exists a scalar $o_{k,3} \geq 0 $ such that (\ref{cons for imperfect ru}) is equivalent to the following LMI
\begin{equation}
     \left[ \begin{matrix}
o_{k,3}\mathbf{I}+\mathbf{M} & \mathbf{\hat{h}}_k^H\mathbf{M} \\
\\\mathbf{M}^H \mathbf{\hat{h}}_k & -\epsilon_k^2-(2^{R_c}-1)\sigma_k^2+\mathbf{\hat{h}}_k^H\mathbf{M}\mathbf{\hat{h}}_k
\end{matrix} \right]  \succeq \mathbf{0}. \label{cons for C c}
\end{equation}
Therefore, the final dual-functional beamforming design problem under imperfect CSI can be reformulated as a tractable convex problem, which is given by
\begin{subequations}
\begin{align}
\max_{ \{\mathbf{W}_i\}_{i=1}^{K+1},\{o_{k,1},o_{k,2},o_{k,3}\}_{k=1}^K}& \quad t_k \label{p: imperfect beam}\\
\text{s.t.} \quad \hat{\Psi}_{k,1}+\hat{\Psi}_{k,2}+\hat{\Psi}_{k,3}&+\tau_{k,3}-\frac{1}{\ln{2}}m_k\tau_{k,4}\geq t_k, \nonumber \\
&\quad \forall k\in \mathcal{K},\forall\theta_e,\forall\phi_e, \label{imperfect t}\\
o_{k,1}\geq 0,o_{k,2} \geq 0,& o_{k,3} \geq 0,\forall k\in\mathcal{K},\\
(\ref{power constraint}),
(\ref{cons for imperfect re}),(\ref{cons for imperfect sensing}),(\ref{cons for S o1})&,(\ref{cons for S o2}),(\ref{cons:Psi1lowerbound})-(\ref{cons:Psi3lowerbound}),(\ref{cons for C c}). \nonumber
\end{align}
\end{subequations}
Furthermore, the auxiliary parameters $\{m_k\}_{k=1}^K$ and $l$ can be updated as 
\begin{equation}
    m_k^*=\frac{1}{\tau_{k,4}},l^*=\frac{1}{\beta},\forall k\in\mathcal{K}.\label{update imperfect ml}
\end{equation}

\subsection{Antenna Position Optimization}
\subsubsection{Convex Reformulation of ~$\Xi_{k,1}$ in Constraint (\ref{cons tk})}
$\Xi_{k,1}$ and $\Xi_{k,2}$ in (\ref{cons tk}) are inherently non-convex. To address this, we tackle them sequentially. First, we simplify (\ref{tau1}) as follows
\begin{equation}
    \Gamma_1(\mathbf{p}_k)+\Gamma_2(\mathbf{p}_k)+\Delta \mathbf{h}_k^H(-\sum\limits_{i=1}^{K}\mathbf{W}_i)\Delta \mathbf{h}_k+2^{\tau_{k,1}}-\sigma_k^2\leq 0,
\end{equation}
where $\Gamma_1(\mathbf{p}_k)$ and $\Gamma_2(\mathbf{p}_k)$ are defined as
\begin{align}
    \Gamma_1(\mathbf{p}_k&)=2\Re\{\mathbf{f}_k^H(\mathbf{p}_k)\mathbf{\Sigma}_k^H\mathbf{G}_k(-\sum\limits_{i=1}^{K}\mathbf{W}_i)\Delta\mathbf{h}_k\}\nonumber\\
    &=2\Re\{\mathbf{f}_k^H(\mathbf{p}_k)\mathbf{\Pi}_{k,1}\Delta \mathbf{h}_k\}=2\Re\{\mathbf{\boldsymbol{c}}_k^\top(\mathbf{p}_k)\Delta\mathbf{h}_k\},
\end{align}
\begin{equation}
    \Gamma_2(\mathbf{p}_k)=\hat{\mathbf{h}}_k^H(-\sum\limits_{i=1}^{K}\mathbf{W}_i)\hat{\mathbf{h}}_k=\mathbf{f}_k^H(\mathbf{p}_k)\mathbf{\Pi}_{k,2}\mathbf{f}_k(\mathbf{p}_k).
\end{equation}
Assume $\delta_{k,4}\mathbf{I}\succeq \nabla^2\Gamma_1(\mathbf{p}_k)$, and by leveraging the second-order Taylor expansion at \( \mathbf{p}_k^{(t)} \), we obtain the surrogate upper bound of $\Gamma_1(\mathbf{p}_k)$ as
\begin{align}
    &\Gamma_1(\mathbf{p}_k)\leq 2\Re\{\mathbf{f}_k^H(\mathbf{p}_k)\mathbf{\Pi}_{k,1}\Delta \mathbf{h}_k\}\nonumber\\
    &+2\Re\{\mathbf{z}^H_{k,1}(\mathbf{p}_k)\Delta \mathbf{h}_k\}+\frac{\delta_{k,4}}{2}(\mathbf{p}_k-\mathbf{p}_k^{(t)})^\top(\mathbf{p}_k-\mathbf{p}_k^{(t)}), \label{Gamma1}
\end{align}
where $
\mathbf{z}_{k,1}(\mathbf{p}_k)=[(\nabla_x \Gamma_1(\mathbf{p}_k^{(t)}))^*~(\nabla_y \Gamma_1(\mathbf{p}_k^{(t)}))^*](\mathbf{p}_k-\mathbf{p}_k^{(t)})$. The terms $\nabla_x \Gamma_1(\mathbf{p}_k)$, $\nabla_y \Gamma_1(\mathbf{p}_k)$ and $\delta_{k,4}$ are given in Appendix \ref{Appendix gammas}.

Next, we apply \textbf{Lemma 2} at $\mathbf{p}_k^{(t)}$ to handle \( \Gamma_2(\mathbf{p}_k) \). 
\begin{align}
    \Gamma_2(\mathbf{p}_k)\leq&M(\mathbf{p}_k)\triangleq\lambda_{\max}(\mathbf{\Pi}_{k,2})L_k\nonumber\\
    &-2\Re\{\mathbf{f}_k^H(\mathbf{p}_k)(\mathbf{\Lambda}_{k,2}-\mathbf{\Pi}_{k,2})\mathbf{f}_k(\mathbf{p}_k^{(t)})\}\nonumber\\
    &+\mathbf{f}_k^H(\mathbf{p}^{(t)}_k)(\mathbf{\Lambda}_{k,2}-\mathbf{\Pi}_{k,2})\mathbf{f}_k(\mathbf{p}_k^{(t)}),\label{Gamma2 part1}
\end{align}
where $\mathbf{\Lambda}_{k,2}=\lambda_{\max}(\mathbf{\Pi}_{k,2})\mathbf{I}$.
Assume $\delta_{k,5}\succeq \nabla^2\Gamma_2(\mathbf{p}_k^{(t)})$, we can get
\begin{align}
   C(\mathbf{p}_k)\leq \hat{C}(\mathbf{p}_k&|\mathbf{p}_k^{(t)})= C(\mathbf{p}^{(t)}_k)+\nabla C(\mathbf{p}^{(t)}_k)(\mathbf{p}_k-\mathbf{p}_k^{(t)})\nonumber\\
   &+\frac{\delta_{k,5}}{2}(\mathbf{p}_k-\mathbf{p}_k^{(t)})^\top(\mathbf{p}_k-\mathbf{p}_k^{(t)}),\label{Gamma2 part2}
\end{align}
where $\nabla C(\mathbf{p}_k^{(t)})$ and $\delta_{k,5}$ are given in the appendix \ref{Appendix Gamma2 part2}. 

Substituting (\ref{Gamma1}), (\ref{Gamma2 part1}), and (\ref{Gamma2 part2}) into (\ref{tau1}), we obtain
\begin{align}
    \Delta \mathbf{h}_k^H(-&\sum\limits_{i=1}^{K}\mathbf{W}_i)\Delta \mathbf{h}_k+\Gamma_1(\mathbf{p}_k)+\Gamma_2(\mathbf{p}_k)+2^{\tau_{k,3}}-\sigma_k^2\nonumber\\
     &\leq\Delta \mathbf{h}_k^H(-\sum\limits_{i=1}^{K}\mathbf{W}_i)\Delta \mathbf{h}_k+2\Re\{(\mathbf{f}_k^H(\mathbf{p}^{(t)}_k)\mathbf{\Pi}_{k,1}
    \nonumber\\
    &+\mathbf{z}_{k,1}^H(\mathbf{p}_k))\Delta\mathbf{h}_k\}+\mu_k(\mathbf{p}_k|\mathbf{p}_k^{(t)})\leq 0, \label{reformulated tau cons}
\end{align}
where $\mu_k(\mathbf{p}_k|\mathbf{p}_k^{(t)})=\frac{\delta_{k,4}}{2}(\mathbf{p}_k-\mathbf{p}_k^{(t)})^\top(\mathbf{p}_k-\mathbf{p}_k^{(t)})+\hat{M}(\mathbf{p}_k|\mathbf{p}_k^{(t)})+2^{\tau_{k,3}}-\sigma_k^2$.
Since $\mu_k(\mathbf{p}_k|\mathbf{p}_k^{(t)})$ is a concave function, the S-procedure cannot be used directly. Instead, we employ \textbf{Lemma 4} to transform it into a tractable form.

\noindent \textbf{Lemma 4} \cite{lemma4,boyd2004convex}: \textit{For any complex Hermitian matrix \( \mathbf{A} \in \mathbb{C}^{N \times N} \), \( \mathbf{b} \in \mathbb{C}^{N \times 1} \), \( c \in \mathbb{R} \), and \( d \in \mathbb{R} \), it follows that}
\begin{equation}
    \begin{bmatrix}
    \mathbf{A} & \mathbf{b} \\
    \mathbf{b}^H & c
    \end{bmatrix} \succeq \mathbf{0} \quad \Rightarrow \quad
    \begin{bmatrix}
    \mathbf{A} & \mathbf{b} \\
    \mathbf{b}^H & d
    \end{bmatrix} \succeq \mathbf{0},
\end{equation}
\textit{holds if and only if \( d \geq c \).}

\textit{Proof: Refer to \cite{lemma4}.}

Subsequently, we introduce an auxiliary variable $\hat{\mu}_k$, satisfying the following inequality
\begin{equation}
    \mu_k(\mathbf{p}_k|\mathbf{p}_k^{(t)})\geq \hat{\mu}_k\label{cons for hat mu}.
\end{equation}
Applying the S-procedure to (\ref{reformulated tau cons}), we assume the existence of a scalar $o_{k,4}\geq 0$ such that (\ref{reformulated tau cons}) is equivalent to the following LMI
\begin{equation}
\scalebox{0.76}{$
    o_{k,4} \left[ \begin{matrix}
\mathbf{I} & \mathbf{0} \\
\mathbf{0} & -\epsilon_k^2
\end{matrix} \right]-
\left[ \begin{matrix}
-\sum\limits_{i=1}^{K}\mathbf{W}_i & \mathbf{\Pi}_{k,1}^{H}\mathbf{f}_k(\mathbf{p}^{(t)}_k)
    +\mathbf{z}_{k,1}(\mathbf{p}_k)\\
\mathbf{f}_k^H(\mathbf{p}^{(t)}_k)\mathbf{\Pi}_{k,1}
    +\mathbf{z}_{k,1}^H(\mathbf{p}_k) & \hat{\mu}_k
\end{matrix} \right] \succeq \mathbf{0}. \label{cons for S antenna tau1}
$}
\end{equation}
\subsubsection{Convex Reformulation of $\Xi_{k,2}$ in Constraint (\ref{cons tk})}
When dealing with $\Xi_{k,2}$, we can further expand (\ref{tau2 cons}) to the following form.
\begin{equation}
    U_{k,1}+U_{k,2}+\Delta \mathbf{h}_k^H\mathbf{V}_k\Delta \mathbf{h}_k+\sigma_k^2-\tau_{k,4}\leq 0, \label{original cons for UK12}
\end{equation}
where
\begin{equation}
    \mathbf{V}_k=\sum\limits_{i=1,i\neq k}^{K}\mathbf{W}_i,
\end{equation}
\begin{equation}
    \mathbf{B}_k=\mathbf{\Sigma}_k^H\mathbf{G}_k\mathbf{V}_k,
\end{equation}
\begin{equation}
    U_{k,1}=2\Re\{\hat{\mathbf{h}}_k^H\mathbf{V}_k\Delta \mathbf{h}_k\}=2\Re\{\mathbf{f}_k^H(\mathbf{p}_k)\mathbf{B}_k\Delta \mathbf{h}\},
\end{equation}
\begin{equation}
    U_{k,2}=\hat{\mathbf{h}}_k^H\mathbf{V}_k\hat{\mathbf{h}}_k.
\end{equation}
Selecting $\delta_{k,6}\mathbf{I}\succeq U_{k,1}(\mathbf{p}_k)$, we can derive an upper bound for $U_{k,1}$ as follows.
\begin{align}
    U_{k,1}&\leq 2\Re\{2\mathbf{f}_k^H(\mathbf{p}_k)\mathbf{B}_k\Delta\mathbf{h}_k\}+[\nabla_xU_{k,1}\nabla_yU_{k,1}](\mathbf{p}_k-\mathbf{p}_k^{(t)})\nonumber\\
    &+\frac{\delta_{k,6}}{2}(\mathbf{p}_k-\mathbf{p}_k^{(t)})^\top(\mathbf{p}_k-\mathbf{p}_k^{(t)})\nonumber \\
    &=2\Re\{\mathbf{f}_k^H(\mathbf{p}_k)\mathbf{B}_k\Delta\mathbf{h}_k\}+2\Re\{\Delta \mathbf{h}_k^H\mathbf{z}_{k,2}\}\nonumber\\
    &+\frac{\delta_{k,6}}{2}(\mathbf{p}_k-\mathbf{p}_k^{(t)})^\top(\mathbf{p}_k-\mathbf{p}_k^{(t)}),\label{Uk,1 uppper}
\end{align}
where $\mathbf{z}_{k,2}$ and $\delta_{k,6}$ are defined in Appendix \ref{Appendix Uk1}. Similarly, for $U_{k,2}$, we have
\begin{align}
U_{k,2}\leq\lambda_{\max}(\mathbf{V}_k)L_k+2\Re\{\mathbf{f}_k^H\mathbf{b}_k\}-\mathbf{f}_k^H(\mathbf{p}_k)\mathbf{b}_k,
\end{align}
where $\mathbf{\Lambda}_{k,3}=\lambda_{\max}(\mathbf{V}_k)\mathbf{I}$ and $\mathbf{b}_k=(\mathbf{\Lambda}_{k,3}-\mathbf{V}_k)\mathbf{f}_k(\mathbf{p}^{(t)})$. Assume $\delta_{k,7}\mathbf{I}\succeq U_{k,2}(\mathbf{p}_k)$ and apply the second-order Taylor expansion at $\mathbf{p}_k^{(t)}$, we obtain
\begin{align}
    &U_{k,2}\leq U_{k,2}(\mathbf{p}_k)+\nabla U_{k,2}(\mathbf{p}_k^{(t)})(\mathbf{p}_k-\mathbf{p}_k^{(t)})\nonumber\\
    &+\delta_{k,7}(\mathbf{p}_k-\mathbf{p}_k^{(t)})^\top(\mathbf{p}_k-\mathbf{p}_k^{(t)}),\label{Uk2 upper bound}
\end{align}
where $\nabla U_{k,2}(\mathbf{p}_k^{(t)})$ and $\delta_{k,7}$ are provided in the Appendix \ref{Appendix uk2}. 

Therefore, substituting (\ref{Uk,1 uppper}) and (\ref{Uk2 upper bound}) into (\ref{original cons for UK12}), we have
\begin{align}
     U_{k,1}+U_{k,2}+\Delta \mathbf{h}_k^H\mathbf{V}_k\Delta &\mathbf{h}_k+\sigma_k^2\leq\Delta \mathbf{h}_k^H\mathbf{V}_k\Delta \mathbf{h}_k\nonumber\\
     +2\Re\{[\mathbf{B}^H\mathbf{f}_k(\mathbf{p}_k^{(t)})+\mathbf{z}_{k,1}]^H&\Delta \mathbf{h}_k\}+\nu_k(\mathbf{p}_k|\mathbf{p}_k^{(t)})    \leq 0,
\end{align}
To apply \textbf{Lemma 4}, we define $\hat{\nu}_k$ as the lower bound for $nu_k(\mathbf{p}_k|\mathbf{p}_k^{(t)})$, which satisfies
\begin{align}
\nu_k(\mathbf{p}_k|&\mathbf{p}_k^{(t)})=\frac{\delta^6_{k}+\delta^7_{k}}{2}(\mathbf{p}_k-\mathbf{p}_k^{(t)})^\top(\mathbf{p}_k-\mathbf{p}_k^{(t)})+\sigma_k^2-\tau_{k,4}\nonumber\\
     &+\nabla T_{k,2}(\mathbf{p}_k^{(t)})(\mathbf{p}_k-\mathbf{p}_k^{(t)})+T_{k,2}(\mathbf{p}_k^{(t)})\geq \hat{\nu}_k, \label{cons for hat nu}
\end{align}
By applying the S-procedure and \textbf{Lemma 4}, there exists a scalar $ o_{k,5}\geq 0$ such that the constraint (\ref{cons for hat nu}) is equivalent to the following LMI.
\begin{equation}
\scalebox{0.93}{$
    o_{k,5} \left[ \begin{matrix}
\mathbf{I} & \mathbf{0} \\
\mathbf{0} & -\epsilon_k^2
\end{matrix} \right]-\left[ \begin{matrix}
-\sum\limits_{i=1,i\neq k}^{K}\mathbf{W}_i & \mathbf{B}_k^H\mathbf{f}_k(\mathbf{p}_k^{(t)})+\mathbf{z}_{k,1}      \\
\mathbf{f}^H_k(\mathbf{p}_k^H)\mathbf{B}_k+\mathbf{z}_{k,1}^H & \hat{\nu}_k
\end{matrix} \right] \succeq \mathbf{0}. \label{cons for S antenna tau2}
$}
\end{equation}

\subsubsection{Convex Reformulation of Constraint (\ref{imperfect ru})}
Let
\begin{align}
    A_k&=2\Re\{\hat{\mathbf{h}}_k^H\mathbf{M}\Delta \mathbf{h}_k\}=2\Re\{\mathbf{f}_k^H(\mathbf{p}_k)\mathbf{\Sigma}_k^H\mathbf{G}_k\mathbf{M}\Delta \mathbf{h}_k\}\nonumber\\
    &=2\Re\{\mathbf{f}_k^H(\mathbf{p}_k)\mathbf{S}_k\Delta \mathbf{h}_k\}=2\Re\{\boldsymbol{\gamma}_k^H\Delta \mathbf{h}_k\}.
\end{align}
Assume $\delta_{k,8}\mathbf{I}\succeq A(\mathbf{p}_k^{(t)})$, thus we have
\begin{equation}
    A_k(\mathbf{p}_k)\leq A_k(\mathbf{p}^{(t)}_k)+\nabla A(\mathbf{p}_k^{(t)})(\mathbf{p}_k-\mathbf{p}^{(t)}_k)+\frac{\delta_{k,8}}{2}(\mathbf{p}_k-\mathbf{p}_k^{(t)}),
\end{equation}
where 
\begin{equation}
    \nabla \mathbf{A}(\mathbf{p}_k)=[\nabla_x\boldsymbol{\gamma}_k\Delta \mathbf{h}_k ~~\nabla_y\boldsymbol{\gamma}_k\Delta \mathbf{h}_k]^\top,
\end{equation}
\begin{equation}
    \delta_{k,8}=\frac{16\pi^2}{\lambda^2}\sum\limits_{l=1}^{l_k}\epsilon_k\sqrt{\mathbf{S}_k[l,:]^H\boldsymbol{S}_k[l,:]}.
\end{equation}
According to (\ref{D upper}), we can 
get the upper bound for $\hat{\mathbf{h}}_k^H\mathbf{M}\hat{\mathbf{h}}_k$, which is $\hat{D}_k(\mathbf{p}_k|\mathbf{p}_k^{(t)})$.
Thus, we can obtain the final form for $\Xi_{k,2}$ as
\begin{equation}
    \Delta \mathbf{h}_k^H\mathbf{M}_k\mathbf{h}_k+2\Re\{\Delta \mathbf{h}_k^H(\mathbf{M}_k^H\hat{\mathbf{h}}_k^{(t)}+\mathbf{z}_{k,2}(\mathbf{p}_k^{(t)})\}+\alpha_k\leq 0, \label{final form for tk}
\end{equation}
where
\begin{equation}
    \mathbf{z}_{k,2}(\mathbf{p}_k|\mathbf{p}_k^{(t)})=[\nabla_x\boldsymbol{\gamma}_k^*~\nabla_y\boldsymbol{\gamma_k}^*](\mathbf{p}_k-\mathbf{p}_k^{(t)}),
\end{equation}
To apply \textbf{Lemma 4}, we introduce a lower bound parameter \(\hat{\alpha}_k\), which satisfies  
\begin{align}  
\alpha_k(\mathbf{p}_k|\mathbf{p}_k^{(t)}) &= \frac{\delta_{k,8}}{2}(\mathbf{p}_k - \mathbf{p}_k^{(t)})^\top(\mathbf{p}_k - \mathbf{p}_k^{(t)}) \nonumber\\  
&\quad + \hat{D}_k(\mathbf{p}_k|\mathbf{p}_k^{(t)}) + (2^{R_c} - 1)\sigma_k^2 \geq \hat{\alpha}_k. \label{cons for hat alpha} 
\end{align}  
Assume there exists a scalar $o_{k,6} \geq 0$ can make (\ref{final form for tk}) equivalent to
\begin{equation}
\scalebox{0.95}{$
    o_{k,6} \left[ \begin{matrix}
\mathbf{I} & \mathbf{0} \\
\mathbf{0} & -\epsilon_k^2
\end{matrix} \right]-\left[ \begin{matrix}
\mathbf{M} &     \boldsymbol{\varrho}_k^H(\mathbf{p}_k|\mathbf{p}_k^{(t)})  \\
\boldsymbol{\varrho}_k(\mathbf{p}_k|\mathbf{p}_k^{(t)}) & \hat{\alpha}_k
\end{matrix} \right] \succeq \mathbf{0}, \label{cons for S antenna ru}
$}
\end{equation}
\begin{equation}
    \boldsymbol{\varrho}_k(\mathbf{p}_k|\mathbf{p}_k^{(t)})=\mathbf{M}^H\hat{\mathbf{h}}_k+\mathbf{z}_{k,2}(\mathbf{p}_k|\mathbf{p}_k^{(t)}).
\end{equation}

Thus, the final optimization problem for antenna position is consequently transformed into a tractable convex problem as follows. 
\begin{subequations}
\begin{align}
\max_{ \{\mathbf{p}_k\}_{k=1}^K,\{o_{k,4},o_{k,5},o_{k,6},\hat{\mu}_k,\hat{\nu}_k,\hat{\alpha}_k\}_{k=1}^K}& \quad \sum\limits_{k=1}^{K}t_k \label{p: imperfect antenna}\\
 \text{s.t.} \quad o_{k,4}\geq 0, o_{k,5}\geq 0, o_{k,6}\geq 0, &\forall k\in\mathcal{K},\\
 (\ref{c5}),(\ref{imperfect t}), (\ref{cons for hat mu}),(\ref{cons for S antenna tau1}),(\ref{cons for hat nu}),(\ref{cons for S antenna tau2})&,(\ref{cons for hat alpha}), (\ref{cons for S antenna ru}). \nonumber
\end{align}
\end{subequations}
Consequently, the algorithm under the imperfect case is summarized in \textbf{Algorithm \ref{algo2}}.

\begin{algorithm}[t]
    \renewcommand{\algorithmicrequire}{\textbf{Initialization:}}
	\renewcommand{\algorithmicensure}{\textbf{Output:}}
    \caption{AO for Solving Problem (\text{P4}).}
     \label{algo2}
    \begin{algorithmic}[1]
        \REQUIRE Set the convergence tolerance $\zeta_1=\zeta_2=\zeta_3=10^{-4}$, the outer iteration index $N_1=0$, beam iteration index $N_2=0$ and FA iteration index $N_3=0$. Initialize $\{\mathbf{p}_k^{(0)}\}_{k=1}^K$ and auxiliary parameters $\{m^{(0)}\}_{k=1}^K$, $l^{(0)}$, maximum iteration index $N_1^{\max}$, $N_2^{\max}$ and $N_1^{\max}$.
        \REPEAT
            \REPEAT 
                \STATE Update $\{\mathbf{W}_k^{(N_2+1)}\}_{k=1}^{K+1}$ by solving (\ref{p: imperfect beam}).
                \STATE Update $\{m_k^{(N_2+1)}\}_{k=1}^K$ and $l^{(N_2+1)}$ by (\ref{update imperfect ml}).
                \STATE Update iteration index $N_1\xleftarrow{}N_1+1$.
            \UNTIL{ the increase of the objective value in (\ref{p: imperfect beam}) is  below the threshold $\zeta_2$ or $N_1>N_1^{\max}$}.
            
            \REPEAT 
                \STATE Update $\{\mathbf{p}_k^{(N_3+1)}\}_{k=1}^K$ by solving (\ref{p: imperfect antenna}).
                \STATE Update iteration index $N_2\xleftarrow{}N_2+1$.
            \UNTIL{the increase of the objective value in (\ref{p: imperfect antenna}) is  below the threshold $\tau$ or $N_2>N_2^{\max}$}. 
             \STATE Update iteration index $N_3\xleftarrow{}N_3+1$, $N_1\xleftarrow{}0$, $N_2\xleftarrow{}0$.
        \UNTIL{the $|\hat{R}^{(N_3)}_{\text{Secure}}-\hat{R}^{(N_3-1)}_{\text{Secure}}| <= \zeta_3$ or $N_3>N_3^{\max}$.}
        \ENSURE $\hat{R}^{(N_3)}_{\text{Secure}}$ and $\mathbf{W}_k^*=\mathbf{w}^*_k\mathbf{w}^{*H}_k$.
    \end{algorithmic}
\end{algorithm}
\subsection{Overall Solution}
Convergence Analysis: In \textbf{Algorithm \ref{algo2}}, each subproblem is addressed locally, which ensures that the objective function (\ref{perfect transformed objective}) does not decrease across iterations. Furthermore, due to the upper bound imposed by the finite transmit power budget on (\ref{perfect transformed objective}), the solution obtained by \textbf{Algorithm \ref{algo2}} is ensured to converge.

Computational Complexity: The computational complexity is mainly dominated by solving (\ref{p: imperfect beam}) and (\ref{p: imperfect antenna}). $\mathcal{O}\big(N_3N_1[(N_T+1)^{3.5}KN_{e,1}N_{e,2}+(KN_{e,1}N_{e,2})^2(N_T+1)^{2.5}+(KN_{e,1}N_{e,2})^3(N_T+1)\ln(\frac{1}{\zeta_1})]$  $+N_3N_2[KL_k^3+N_T^{1.5}\ln(\frac{1}{\zeta_2})] \big)$.

%% file: Chapter/simulation.tex
\section{Simulation Results}
In this section, we present simulation results to validate the secure transmission performance in the FA-aided RSMA-ISAC system. In this study, users are randomly positioned around the BS. The distance $d_k$ between the $k^{th}$ user and the BS is randomly distributed within the range of $[20 \text{m}, 100 \text{m}]$. Meanwhile, the distance $d_E$ between the eavesdropping target and the BS follows a uniform distribution, i.e., $d_E\sim U[10 \text{m}, 15 \text{m}]$. A carrier frequency of 3 GHz is employed, corresponding to a carrier wavelength of $\lambda = 0.1$ m. For the communication channel, the path numbers are $L_1=...=L_k=12$. Moreover, the PRM $\boldsymbol{\Sigma}_k=\text{diag}\{\sigma_{k,1},\sigma_{k,2},...,\sigma_{k,L_k}\}$, the elements of the PRM follow a circularly symmetric complex Gaussian (CSCG) distribution, i.e., $\sigma_{k,l} \sim \mathcal{CN}\left(0, \frac{C_{k}}{L_k}\right)$. Here, $C_{k} = C_0 d_k^{-\tau}$ represents the large-scale path loss, where $C_0 = \left(\frac{\lambda}{4\pi^2}\right)^2$ is the expected average channel power gain at the reference distance of 1 m. The path-loss exponent is $\tau = 2.8$, and $d_k$ represents the distance between the $k^{th}$ user and the BS. Similarly, the large-scale path loss for the sensing channel is assumed as $\eta_e = C_0 d_E^{-\tau}$. The received noise power for each user and Eve is assumed to be $-80$ dBm. The parameters $N_{e,1}=2$, $N_{e,2}=2$ and the $\vartheta_e\in\{\frac{\pi}{10},\frac{\pi}{4}\}$ and $\vartheta_e\in\{\frac{\pi}{10},\frac{\pi}{3}\}$ are used in the simulation. The communication error is defined as $\hat{\epsilon}_k=\frac{\epsilon_k^2}{|\mathbf{h}_k|}$. The feasible region of each LU is modeled as a 2D square, i.e., $\mathcal{C}_k=\mathcal{C}=[-\frac{A_0}{2},\frac{A_0}{2}]\times [-\frac{A_0}{2},\frac{A_0}{2}]$. For simplicity, $\alpha_k=\frac{1}{K}, \forall k\in \mathcal{K}$, the predefined values $R_c\in[0.5,2]$ and $S_0=[0.02,2]*10^{-6}$ increase with the transmit power, which varies from 20 dBm to 40 dBm.

In addition to the proposed FA-based algorithm \textbf{(Prop-FA)}, we also consider the conventional SDMA method, which can be easily obtained by setting $R_c=0$. The following baseline schemes are included below for comparison.

\label{simulation}
\begin{itemize}
    \item \textbf{FA-SDMA}: The FA positions of $K$ LUs are obtained by solving the same optimization problems but setting $R_c=0$.
    
    \item \textbf{FPA-RSMA}: Each user's FA is fixed at the origin. The dual-functional beamforming is obtained in the same way as Prop-FA.
    
    \item \textbf{FPA-SDMA}: The dual-functional beamforming is obtained in the same way as Prop-FA with fixed antenna positions but setting $R_c=0$.
\end{itemize}

\begin{figure}[t]
    \centering
    \includegraphics[width=0.9\linewidth]{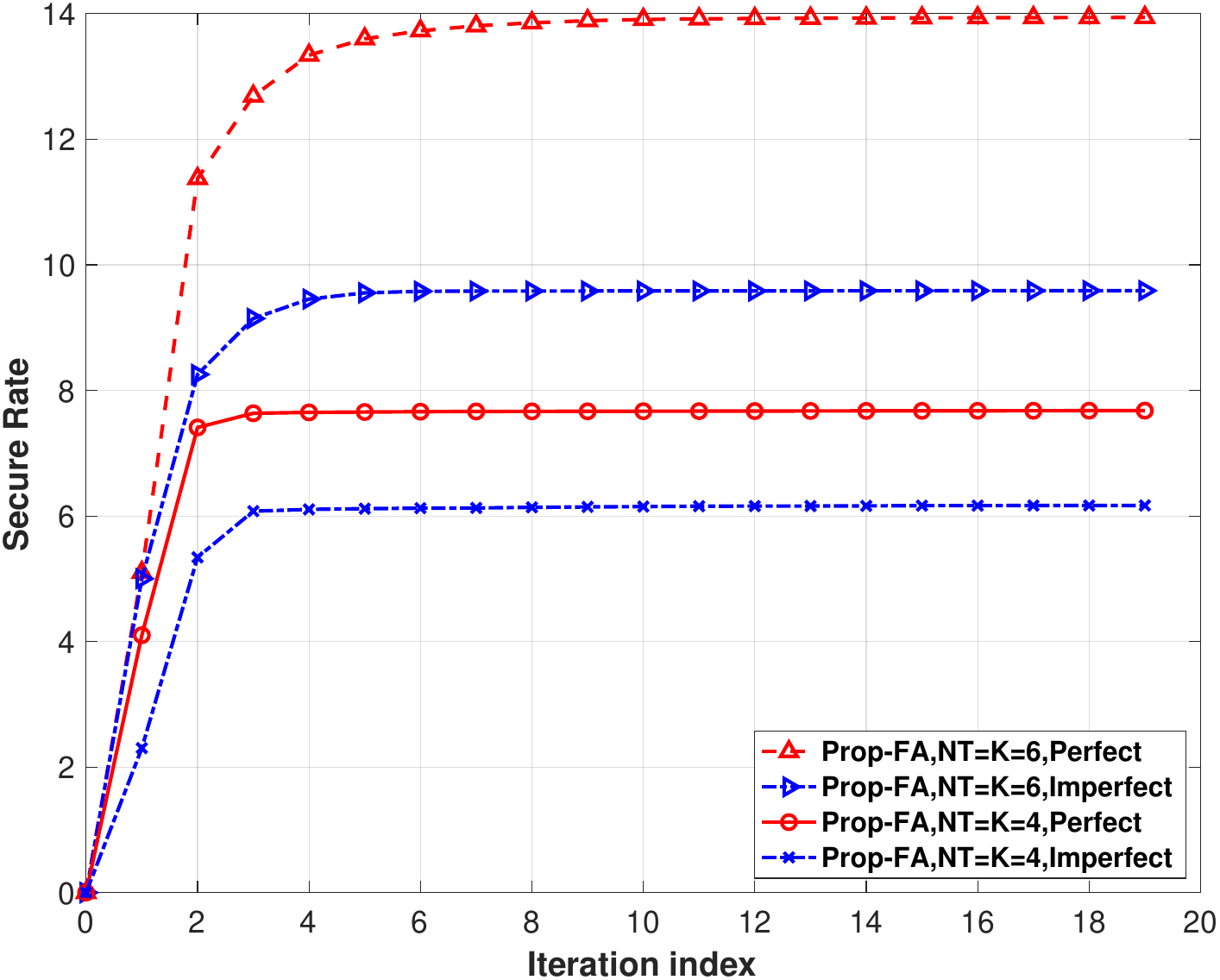}
    \caption{The convergence behavior under perfect and imperfect cases. $A_0=3\lambda$.}
    \label{fig: Convergence analysis.}
\end{figure}

Fig. \ref{fig: Convergence analysis.} illustrates the convergence of \textbf{Algorithm \ref{algo1}} and \textbf{Algorithm \ref{algo2}} with varying numbers of FAs under both perfect and imperfect conditions. As \(N_T\) and \(K\) increase, the algorithms require more iterations to converge. However, it can be observed that regardless of whether the conditions are perfect or imperfect, the algorithms still converge within a relatively small number of iterations. Specifically, in the ideal case where \(N_T = K = 6\), the algorithms converge in approximately 10 iterations, while under imperfect conditions, convergence occurs in about 8 iterations. The convergence behavior shown in the figure is consistent with the earlier discussions.

\begin{figure}[t]
    \centering
    \includegraphics[width=0.88\linewidth]{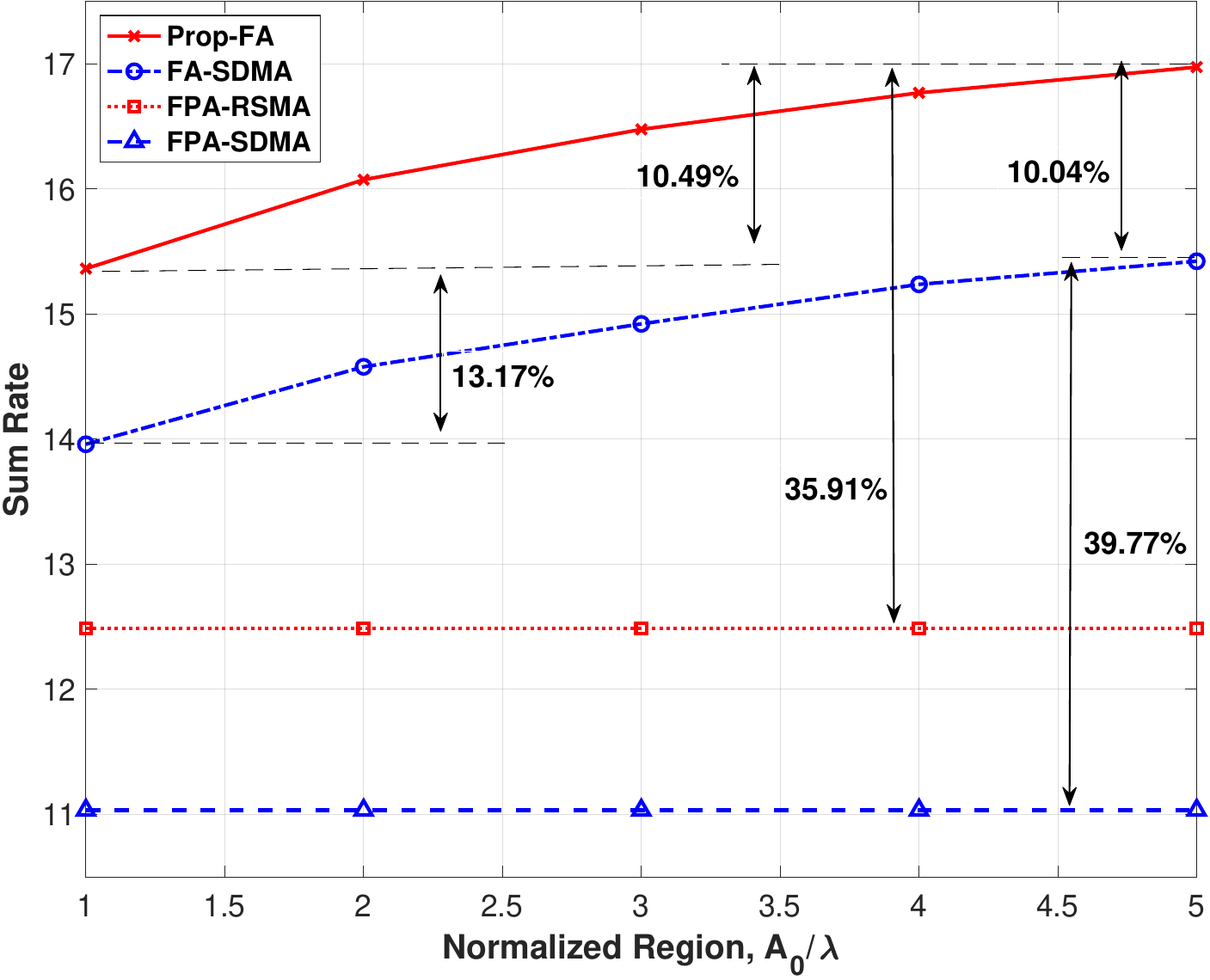}
    \caption{Secure sum-rate versus the normalized region size with perfect CSI. $P_0=30\text{dBm}$, $N_T=K=4$.}
    \label{fig1: perfect normalized region size}
\end{figure}

Fig. \ref{fig1: perfect normalized region size} illustrates the performance of the proposed algorithm and baseline schemes under perfect CSI. On the one hand, the flexible adjustment of FA positions significantly enhances performance. Specifically, the proposed algorithm \textbf{Prop-FA} achieves a performance improvement of 10.49\% when $A_0$ increases from \(1\lambda\) to \(5\lambda\), demonstrating its capability to utilize a larger feasible region under ideal conditions effectively. Both SDMA and RSMA schemes benefit from the proposed FA positions optimization, with RSMA achieving a maximum gain of 35.91\% and SDMA reaching up to 39.77\%. On the other hand, RSMA consistently outperforms SDMA in all scenarios. Specifically, when FPAs are employed, the proposed algorithm \textbf{Prop-FA} achieves a performance gain of 13.17\% over \textbf{FA-SDMA}. When $A_0$ is constrained to \(5\lambda\), this gain adjusts to 10.04\%. These results demonstrate the effectiveness of the proposed algorithm in fully harnessing the combined benefits of FA and RSMA.

\begin{figure}[t]
    \centering
    \includegraphics[width=0.88\linewidth]{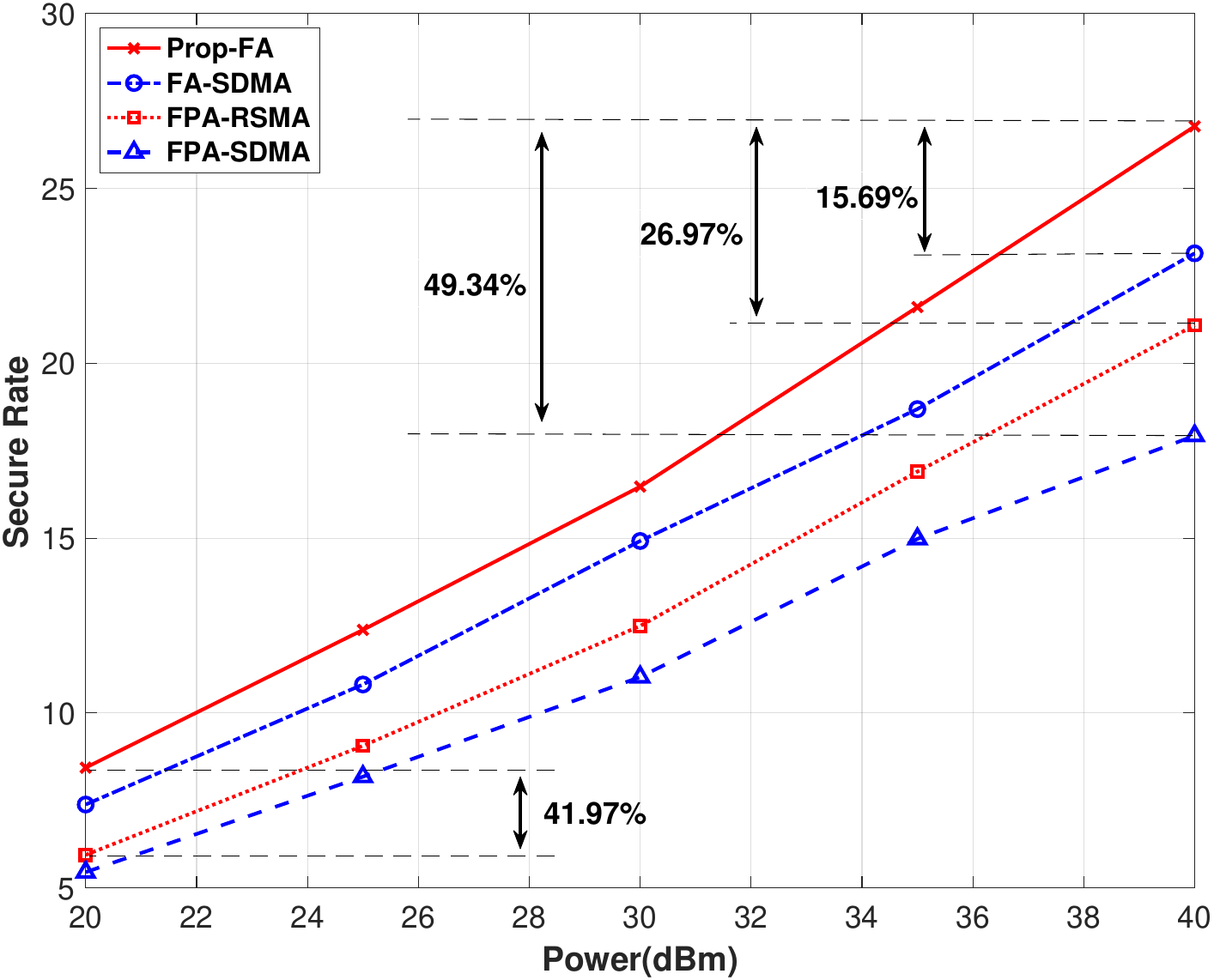}
    \caption{Secure sum-rate versus the power with perfect CSI.  $A_0=3\lambda$, $N_T=K=4$.}
    \label{fig: perfect different power}
\end{figure}

Fig. \ref{fig: perfect different power} demonstrates the performance comparison of various schemes under different power levels in perfect CSI scenarios. The results reveal that the proposed method with FA position optimization consistently delivers substantial performance gains, regardless of whether the power is low or high. Specifically, at $P =40$ dBm, the proposed method achieves gains of 15.69\%, 26.97\%, and 49.34\% over \textbf{FA-SDMA}, \textbf{FPA-RSMA}, and \textbf{FPA-SDMA}, respectively. When $P =20$ dBm, FA positions optimization provides a significant gain of up to 41.97\% compared to \textbf{FPA-RSMA}. These findings underscore the adaptability of the proposed method to varying power conditions.

\begin{figure}[t]
    \centering
    \includegraphics[width=0.88\linewidth]{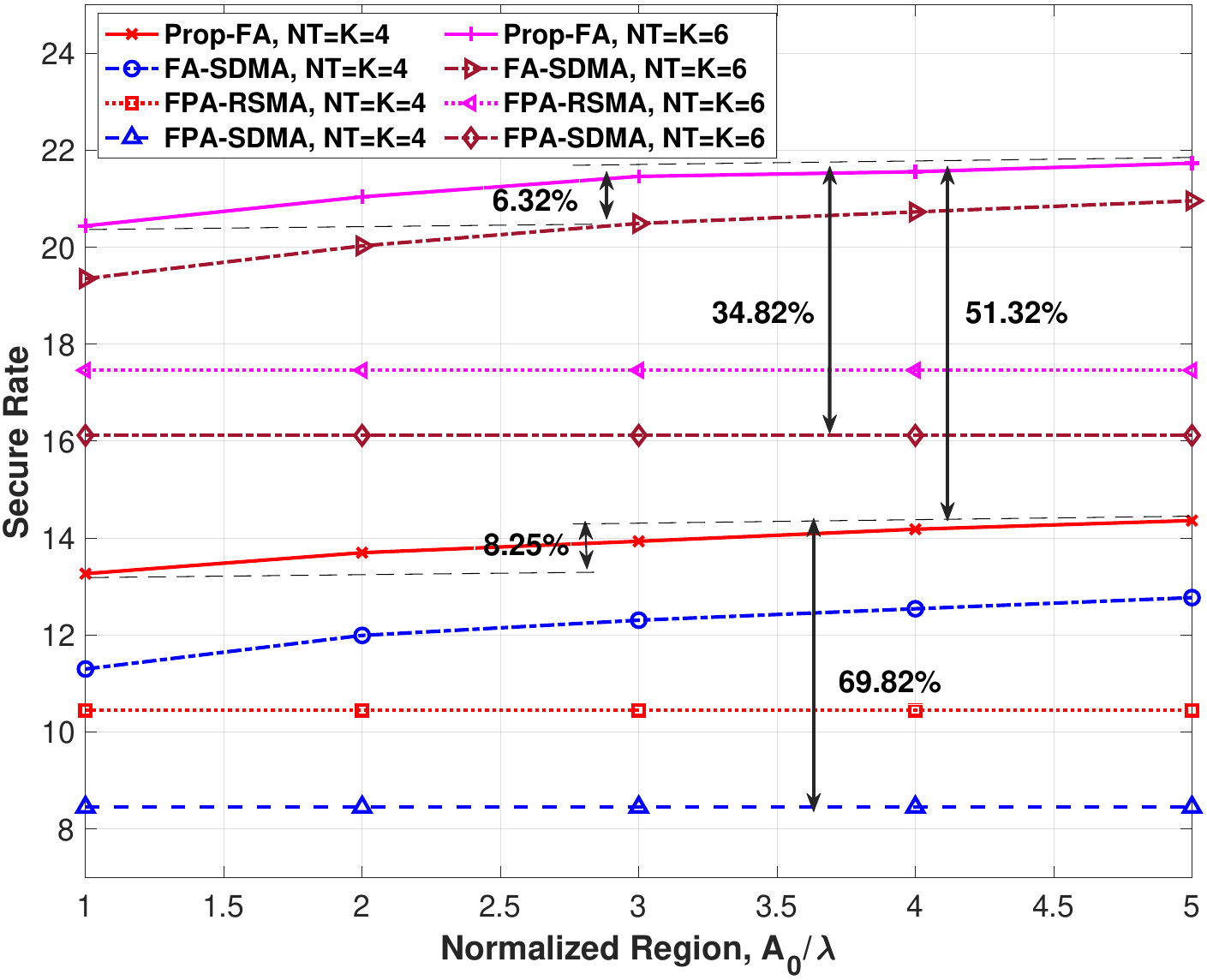}
    \caption{Secure sum-rate versus normalized region size with imperfect CSI. $P_0=30\text{dBm}$, $\hat{\epsilon}_k=0.01$.}
    \label{fig: imperfect different range}
\end{figure}

Fig. \ref{fig: imperfect different range} compares the proposed algorithm and baseline methods under imperfect CSI conditions with varying user and antenna configurations across different feasible region sizes. As the number of antennas and LUs increases, the secure rate exhibits a corresponding rise. For instance, when the number of users \(K\) increases from 4 to 6, the proposed approach achieves a performance improvement of 51.32\%. In scenarios with \(K = 4\), the benefits of optimizing antenna positions are more pronounced compared to cases with a larger number of LUs. Specifically, with a feasible region $A_0$ of \(5\lambda\), the performance improves by 8.25\% when $A_0$ expands from \(1\lambda\) to \(5\lambda\) for \(K = 4\), while a gain of 6.32\% is observed for \(K = 6\). This demonstrates the diminishing marginal effect of FA optimization: as the number of users grows, the performance gain from the same feasible region expansion gradually decreases. Nevertheless, even under imperfect CSI conditions, the proposed algorithm consistently delivers significant performance improvements compared to the traditional \textbf{FPA-SDMA} across diverse user and antenna configurations. For example, with 4 LUs, the proposed method achieves a gain of up to 69.82\%, while maintaining a gain of 34.82\% for \(K = 6\).

\begin{figure}[t]
    \centering
    \includegraphics[width=0.88\linewidth]{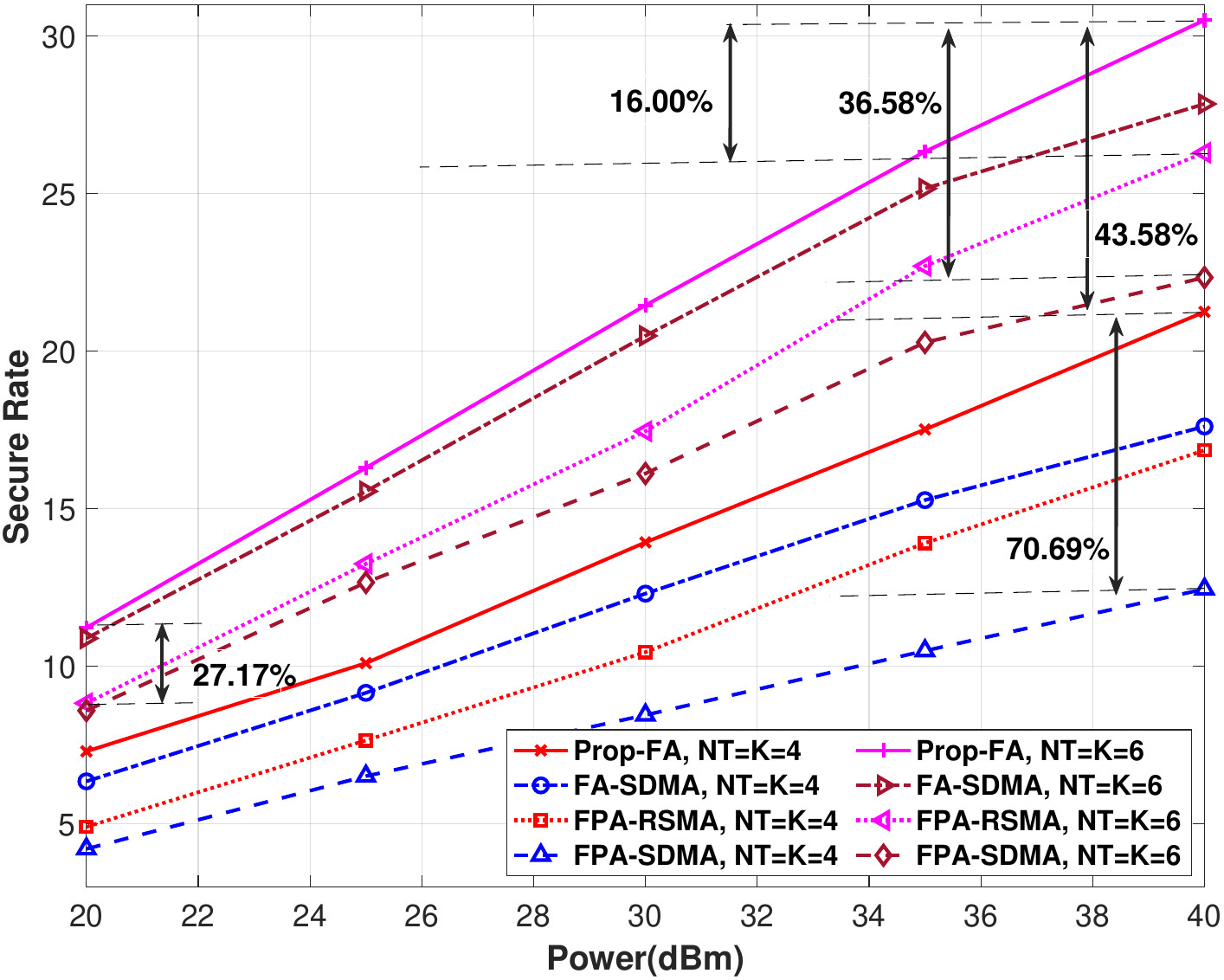}
    \caption{Secure sum-rate versus the power with imperfect CSI. $A_0=3\lambda$, $\hat{\epsilon}_k=0.01$.}
    \label{fig: imperfect different power}
\end{figure}

Fig. \ref{fig: imperfect different power} depicts the variation of the secure rate as a function of transmit power under different user and antenna configurations with imperfect CSI. On the one hand, at low power levels, RSMA and SDMA exhibit comparable performance for \(K = 6\); however, as the power increases, the DoF advantage of RSMA becomes apparent. On the other hand, the proposed FA position optimization achieves notable gains across different power levels. For example, with \(K = 6\), the proposed method achieves gains of 27.17\% and 16.00\% over \textbf{FPA-RSMA} at \(P = 20\) dBm and \(P = 40\) dBm, respectively. Compared to the conventional \textbf{FPA-SDMA}, the proposed method delivers gains of up to 70.69\% for \(K = 4\) and 43.58\% for \(K = 6\) at \(P = 40\) dBm. These results demonstrate that the proposed algorithm effectively leverages both spatial DoF and stream DoF to enhance the worst-case secure rate across diverse power levels.

\begin{figure}[t]
    \centering
    \includegraphics[width=0.9\linewidth]{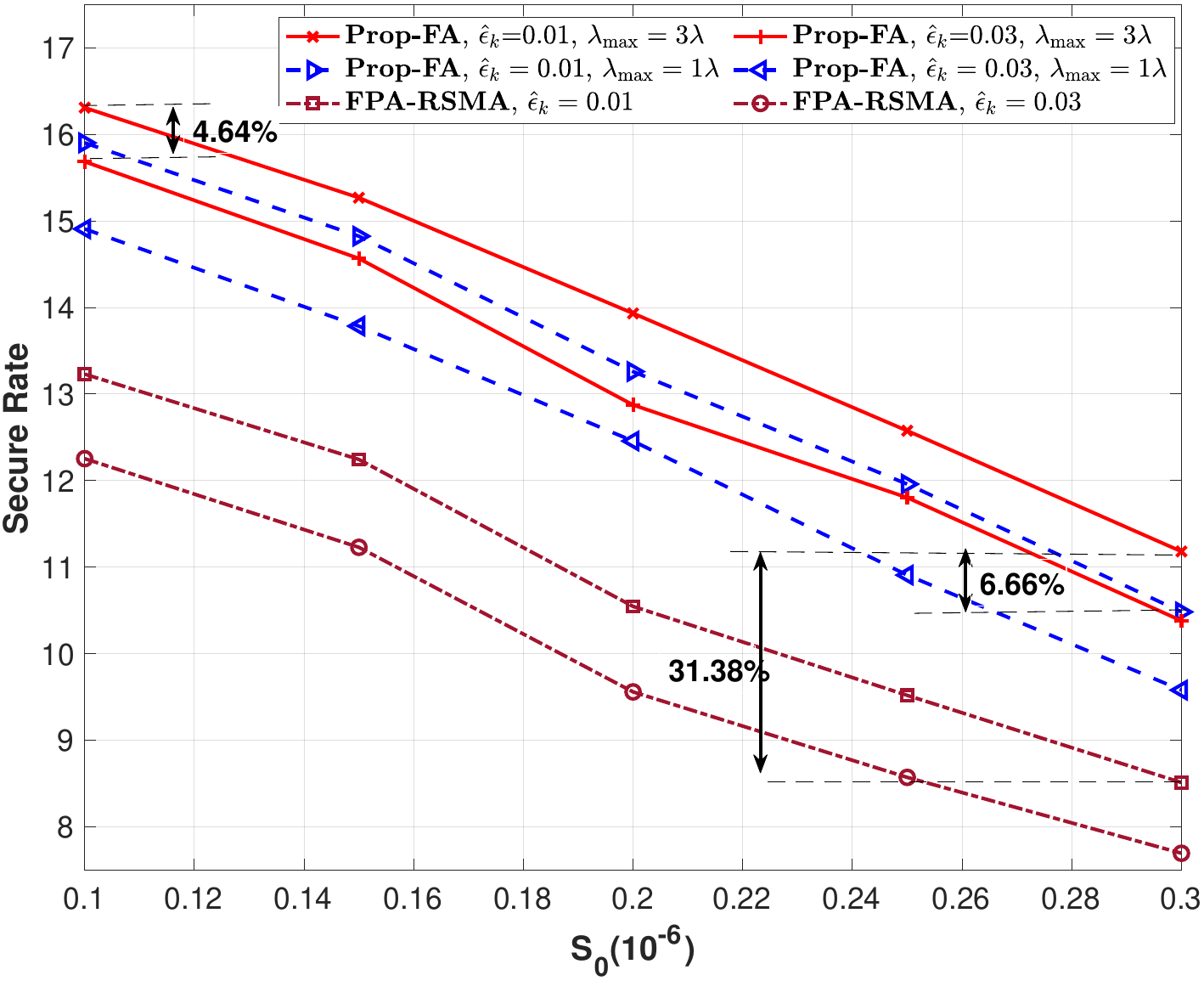}
    \caption{Secure sum-rate versus the $S_0$ with imperfect CSI. $P_0$=30dBm, $N_T=K=4$.}
    \label{fig: imperfect different S0}
\end{figure}

Fig. \ref{fig: imperfect different S0} compares the secure rate of the proposed algorithm \textbf{Prop-FA} under different feasible region sizes and \textbf{RSMA-FPA} scheme across varying sensing thresholds. The results demonstrate that a larger feasible region can better accommodate higher sensing requirements. Specifically, when the sensing requirement reaches its maximum value, i.e., \(S_0 = 0.3 \times 10^{-6}\), \textbf{Prop-FA} with \(A_0 = 3\lambda\) achieves a 6.66\% improvement compared to \(A_0 = 1\lambda\) and a significant 31.38\% improvement over the FPA configuration. Moreover, channel estimation errors have a notable impact on the secure rate; however, a larger feasible region provides better robustness against such errors. For instance, with \(A_0 = 3\lambda\) and \textbf{Prop-FA}, although the performance degrades by 4.64\% when \(\hat{\epsilon}_k\) increases from 0.01 to 0.03, it still significantly outperforms the FPA algorithm even under lower channel estimation errors. The proposed robust FA-based algorithm not only mitigates the adverse effects of imperfect CSI but also effectively meets higher sensing requirements.

%% file: Chapter/conclusion.tex
\section{Conclusion}
In conclusion, this work investigates a novel FA-aided downlink RSMA-ISAC system that, for the first time, fully leverages the flexibility of user streams and the spatial adaptability of FAs. By jointly optimizing the positions of FAs, as well as the common and private beamformers, we aim to maximize the secure sum rate under both perfect and imperfect CSI conditions. We propose an AO-based framework to iteratively optimize the beamforming matrix, auxiliary variables, and FA positions. Simulation results validate the robust synergistic effectiveness of the proposed scheme compared to traditional FPA-based and SDMA-based schemes across different user and antenna configurations, varying channel estimation errors, and diverse sensing requirements. The proposed scheme consistently achieves superior secure performance, highlighting the potential of FA-aided RSMA-ISAC systems as a promising solution for future wireless networks, particularly in scenarios demanding high throughput and enhanced privacy.
\label{conclusion}